\documentclass[longauth]{aa}

\usepackage{graphicx}
\usepackage{natbib}
\usepackage{scalerel}
\usepackage{enumitem}
\usepackage[table]{xcolor}
\usepackage{float}
\usepackage{txfonts}
\usepackage{tikz}
\usetikzlibrary{positioning, arrows.meta}
\usepackage[pdfencoding=auto,psdextra]{hyperref}
\hypersetup{
    colorlinks=true,
    linkcolor=blue,
    filecolor=magenta,      
    urlcolor=blue,
    citecolor=blue
}
\makeatletter
\renewcommand*\aa@pageof{, page \thepage{} of \pageref*{LastPage}}
\makeatother

\usepackage[utf8]{inputenc}

\usepackage[switch, modulo]{lineno}
              
\renewcommand{\linenumbers}[0]{}

\usepackage{euclid}

\newcommand{\ignoreme}[1]{}

\begin{document}
%
%

\title{Euclid Data Release 1 (DR1)}
\subtitle{AstroVink-TREX: Automated strong gravitational lens segmentation as input for mass modelling pipelines}


\newcommand{\orcid}[1]{} 
\author{Euclid Collaboration: S.~H.~Vincken\orcid{0009-0005-7305-2359}\thanks{\email{saamie.vincken@fhnw.ch}}\inst{\ref{aff1},\ref{aff2},\ref{aff3}}
\and N.~E.~P.~Lines\orcid{0009-0004-7751-1914}\inst{\ref{aff1}}
\and J.~W.~Nightingale\orcid{0000-0002-8987-7401}\inst{\ref{aff4}}
\and T.~E.~Collett\orcid{0000-0001-5564-3140}\inst{\ref{aff1}}
\and L.~R.~Ecker\orcid{0009-0005-3508-2469}\inst{\ref{aff5},\ref{aff6},\ref{aff7}}
\and S.~Schuldt\orcid{0000-0003-2497-6334}\inst{\ref{aff8},\ref{aff9},\ref{aff10}}
\and A.~Nersesian\orcid{0000-0001-6843-409X}\inst{\ref{aff11},\ref{aff12}}
\and T.~Li\orcid{0009-0005-5008-0381}\inst{\ref{aff1}}
\and N.~Wielders\orcid{0009-0001-8110-819X}\inst{\ref{aff4}}
\and F.~Courbin\orcid{0000-0003-0758-6510}\inst{\ref{aff13},\ref{aff14},\ref{aff15}}
\and G.~Despali\orcid{0000-0001-6150-4112}\inst{\ref{aff16},\ref{aff17},\ref{aff18}}
\and K.~Rojas\orcid{0000-0003-1391-6854}\inst{\ref{aff19}}
\and C.~J.~Storfer\orcid{0000-0002-0385-0014}\inst{\ref{aff20}}
\and A.~Angress\orcid{0009-0005-5250-9487}\inst{\ref{aff21},\ref{aff22}}
\and D.~J.~Bowden\orcid{0009-0008-6114-1401}\inst{\ref{aff23}}
\and W.~J.~R.~Enzi\orcid{0009-0004-2992-3148}\inst{\ref{aff1},\ref{aff24}}
\and K.~Finner\orcid{0000-0002-4462-0709}\inst{\ref{aff25}}
\and P.~Fleury\orcid{0000-0001-9292-3651}\inst{\ref{aff26}}
\and A.~Galan\orcid{0000-0003-2547-9815}\inst{\ref{aff27}}
\and R.~Gavazzi\orcid{0000-0002-5540-6935}\inst{\ref{aff28},\ref{aff29}}
\and Q.~He\inst{\ref{aff30}}
\and N.~B.~Hogg\orcid{0000-0001-9346-4477}\inst{\ref{aff31},\ref{aff32}}
\and D.~P.~Johnson\orcid{0000-0002-0311-2513}\inst{\ref{aff26}}
\and S.~K.~Li\orcid{0000-0002-4490-7304}\inst{\ref{aff33}}
\and A.~Melo\orcid{0000-0002-6449-3970}\inst{\ref{aff34},\ref{aff35},\ref{aff36}}
\and M.~Meneghetti\orcid{0000-0003-1225-7084}\inst{\ref{aff17},\ref{aff18}}
\and R.~B.~Metcalf\orcid{0000-0003-3167-2574}\inst{\ref{aff16},\ref{aff17}}
\and M.~Millon\orcid{0000-0001-7051-497X}\inst{\ref{aff37}}
\and C.~Olave\inst{\ref{aff38}}
\and C.~M.~O'Riordan\orcid{0000-0003-2227-1998}\inst{\ref{aff7}}
\and H.~Qu\orcid{0009-0006-0299-0265}\inst{\ref{aff25}}
\and C.~Saulder\orcid{0000-0002-0408-5633}\inst{\ref{aff6},\ref{aff5}}
\and D.~Sluse\orcid{0000-0001-6116-2095}\inst{\ref{aff11}}
\and G.~Smadja\inst{\ref{aff39}}
\and A.~Sonnenfeld\orcid{0000-0002-6061-5977}\inst{\ref{aff40},\ref{aff41},\ref{aff42}}
\and H.~Srivastava\orcid{0009-0002-3254-0952}\inst{\ref{aff43}}
\and T.~T.~Thai\orcid{0000-0002-8408-4816}\inst{\ref{aff44}}
\and C.~Tortora\orcid{0000-0001-7958-6531}\inst{\ref{aff45}}
\and Y.~Xie\inst{\ref{aff46},\ref{aff7},\ref{aff47}}
\and B.~Altieri\orcid{0000-0003-3936-0284}\inst{\ref{aff48}}
\and A.~Amara\inst{\ref{aff49}}
\and S.~Andreon\orcid{0000-0002-2041-8784}\inst{\ref{aff50}}
\and N.~Auricchio\orcid{0000-0003-4444-8651}\inst{\ref{aff17}}
\and C.~Baccigalupi\orcid{0000-0002-8211-1630}\inst{\ref{aff51},\ref{aff52},\ref{aff53},\ref{aff54}}
\and M.~Baldi\orcid{0000-0003-4145-1943}\inst{\ref{aff38},\ref{aff17},\ref{aff18}}
\and S.~Bardelli\orcid{0000-0002-8900-0298}\inst{\ref{aff17}}
\and P.~Battaglia\orcid{0000-0002-7337-5909}\inst{\ref{aff17}}
\and F.~Bernardeau\orcid{0009-0007-3015-2581}\inst{\ref{aff55},\ref{aff29}}
\and A.~Biviano\orcid{0000-0002-0857-0732}\inst{\ref{aff52},\ref{aff51}}
\and M.~Bolzonella\orcid{0000-0003-3278-4607}\inst{\ref{aff17}}
\and E.~Branchini\orcid{0000-0002-0808-6908}\inst{\ref{aff56},\ref{aff57},\ref{aff50}}
\and M.~Brescia\orcid{0000-0001-9506-5680}\inst{\ref{aff58},\ref{aff45}}
\and S.~Camera\orcid{0000-0003-3399-3574}\inst{\ref{aff59},\ref{aff60},\ref{aff61}}
\and V.~Capobianco\orcid{0000-0002-3309-7692}\inst{\ref{aff61}}
\and C.~Carbone\orcid{0000-0003-0125-3563}\inst{\ref{aff10}}
\and J.~Carretero\orcid{0000-0002-3130-0204}\inst{\ref{aff62},\ref{aff63}}
\and M.~Castellano\orcid{0000-0001-9875-8263}\inst{\ref{aff64}}
\and G.~Castignani\orcid{0000-0001-6831-0687}\inst{\ref{aff17}}
\and S.~Cavuoti\orcid{0000-0002-3787-4196}\inst{\ref{aff45},\ref{aff65}}
\and A.~Cimatti\inst{\ref{aff66}}
\and C.~Colodro-Conde\inst{\ref{aff67}}
\and G.~Congedo\orcid{0000-0003-2508-0046}\inst{\ref{aff68}}
\and L.~Conversi\orcid{0000-0002-6710-8476}\inst{\ref{aff69},\ref{aff48}}
\and Y.~Copin\orcid{0000-0002-5317-7518}\inst{\ref{aff39}}
\and H.~M.~Courtois\orcid{0000-0003-0509-1776}\inst{\ref{aff70}}
\and M.~Cropper\orcid{0000-0003-4571-9468}\inst{\ref{aff71}}
\and H.~Degaudenzi\orcid{0000-0002-5887-6799}\inst{\ref{aff27}}
\and G.~De~Lucia\orcid{0000-0002-6220-9104}\inst{\ref{aff52}}
\and H.~Dole\orcid{0000-0002-9767-3839}\inst{\ref{aff72}}
\and F.~Dubath\orcid{0000-0002-6533-2810}\inst{\ref{aff27}}
\and X.~Dupac\inst{\ref{aff48}}
\and S.~Escoffier\orcid{0000-0002-2847-7498}\inst{\ref{aff73}}
\and M.~Farina\orcid{0000-0002-3089-7846}\inst{\ref{aff74}}
\and R.~Farinelli\inst{\ref{aff17}}
\and F.~Faustini\orcid{0000-0001-6274-5145}\inst{\ref{aff64}}
\and S.~Ferriol\inst{\ref{aff39}}
\and F.~Finelli\orcid{0000-0002-6694-3269}\inst{\ref{aff17},\ref{aff75}}
\and M.~Frailis\orcid{0000-0002-7400-2135}\inst{\ref{aff52}}
\and E.~Franceschi\orcid{0000-0002-0585-6591}\inst{\ref{aff17}}
\and M.~Fumana\orcid{0000-0001-6787-5950}\inst{\ref{aff10}}
\and L.~Gabarra\orcid{0000-0002-8486-8856}\inst{\ref{aff23}}
\and S.~Galeotta\orcid{0000-0002-3748-5115}\inst{\ref{aff52}}
\and K.~George\orcid{0000-0002-1734-8455}\inst{\ref{aff76}}
\and B.~Gillis\orcid{0000-0002-4478-1270}\inst{\ref{aff68}}
\and C.~Giocoli\orcid{0000-0002-9590-7961}\inst{\ref{aff17},\ref{aff18}}
\and J.~Gracia-Carpio\orcid{0000-0003-4689-3134}\inst{\ref{aff6}}
\and A.~Grazian\orcid{0000-0002-5688-0663}\inst{\ref{aff77}}
\and F.~Grupp\inst{\ref{aff6},\ref{aff5}}
\and W.~G.~Hartley\orcid{0000-0002-4563-5101}\inst{\ref{aff27}}
\and S.~V.~H.~Haugan\orcid{0000-0001-9648-7260}\inst{\ref{aff78}}
\and S.~Hemmati\orcid{0000-0003-2226-5395}\inst{\ref{aff25}}
\and W.~Holmes\orcid{0009-0007-8554-4646}\inst{\ref{aff79}}
\and F.~Hormuth\inst{\ref{aff80}}
\and A.~Hornstrup\orcid{0000-0002-3363-0936}\inst{\ref{aff81},\ref{aff82}}
\and M.~Huertas-Company\orcid{0000-0002-1416-8483}\inst{\ref{aff67},\ref{aff83},\ref{aff84}}
\and K.~Jahnke\orcid{0000-0003-3804-2137}\inst{\ref{aff85}}
\and M.~Jhabvala\inst{\ref{aff86}}
\and S.~Kermiche\orcid{0000-0002-0302-5735}\inst{\ref{aff73}}
\and A.~Kiessling\orcid{0000-0002-2590-1273}\inst{\ref{aff79}}
\and B.~Kubik\orcid{0009-0006-5823-4880}\inst{\ref{aff39}}
\and M.~K\"ummel\orcid{0000-0003-2791-2117}\inst{\ref{aff5}}
\and M.~Kunz\orcid{0000-0002-3052-7394}\inst{\ref{aff37}}
\and H.~Kurki-Suonio\orcid{0000-0002-4618-3063}\inst{\ref{aff9},\ref{aff87}}
\and A.~M.~C.~Le~Brun\orcid{0000-0002-0936-4594}\inst{\ref{aff88}}
\and S.~Ligori\orcid{0000-0003-4172-4606}\inst{\ref{aff61}}
\and P.~B.~Lilje\orcid{0000-0003-4324-7794}\inst{\ref{aff78}}
\and V.~Lindholm\orcid{0000-0003-2317-5471}\inst{\ref{aff9},\ref{aff87}}
\and I.~Lloro\orcid{0000-0001-5966-1434}\inst{\ref{aff89}}
\and M.~Magliocchetti\orcid{0000-0001-9158-4838}\inst{\ref{aff74}}
\and G.~Mainetti\orcid{0000-0003-2384-2377}\inst{\ref{aff90}}
\and O.~Mansutti\orcid{0000-0001-5758-4658}\inst{\ref{aff52}}
\and S.~Marcin\inst{\ref{aff19}}
\and O.~Marggraf\orcid{0000-0001-7242-3852}\inst{\ref{aff91}}
\and M.~Martinelli\orcid{0000-0002-6943-7732}\inst{\ref{aff64},\ref{aff92}}
\and N.~Martinet\orcid{0000-0003-2786-7790}\inst{\ref{aff28}}
\and F.~Marulli\orcid{0000-0002-8850-0303}\inst{\ref{aff16},\ref{aff17},\ref{aff18}}
\and R.~J.~Massey\orcid{0000-0002-6085-3780}\inst{\ref{aff93}}
\and N.~Mauri\orcid{0000-0001-8196-1548}\inst{\ref{aff66},\ref{aff18}}
\and E.~Medinaceli\orcid{0000-0002-4040-7783}\inst{\ref{aff17}}
\and S.~Mei\orcid{0000-0002-2849-559X}\inst{\ref{aff94},\ref{aff95}}
\and M.~Melchior\inst{\ref{aff3}}
\and E.~Merlin\orcid{0000-0001-6870-8900}\inst{\ref{aff77}}
\and G.~Meylan\orcid{0000-0001-6503-0209}\inst{\ref{aff96}}
\and P.~Monaco\orcid{0000-0003-2083-7564}\inst{\ref{aff97},\ref{aff52},\ref{aff53},\ref{aff51}}
\and A.~Mora\orcid{0000-0002-1922-8529}\inst{\ref{aff98}}
\and M.~Moresco\orcid{0000-0002-7616-7136}\inst{\ref{aff16},\ref{aff17}}
\and C.~Moretti\orcid{0000-0003-3314-8936}\inst{\ref{aff52},\ref{aff51},\ref{aff53}}
\and L.~Moscardini\orcid{0000-0002-3473-6716}\inst{\ref{aff16},\ref{aff17},\ref{aff18}}
\and C.~Neissner\orcid{0000-0001-8524-4968}\inst{\ref{aff99},\ref{aff63}}
\and R.~C.~Nichol\orcid{0000-0003-0939-6518}\inst{\ref{aff49}}
\and S.-M.~Niemi\orcid{0009-0005-0247-0086}\inst{\ref{aff100}}
\and C.~Padilla\orcid{0000-0001-7951-0166}\inst{\ref{aff99}}
\and S.~Paltani\orcid{0000-0002-8108-9179}\inst{\ref{aff27}}
\and F.~Pasian\orcid{0000-0002-4869-3227}\inst{\ref{aff52}}
\and W.~J.~Percival\orcid{0000-0002-0644-5727}\inst{\ref{aff101},\ref{aff102},\ref{aff103}}
\and V.~Pettorino\orcid{0000-0002-4203-9320}\inst{\ref{aff100}}
\and A.~Pezzotta\orcid{0000-0003-0726-2268}\inst{\ref{aff50}}
\and S.~Pires\orcid{0000-0002-0249-2104}\inst{\ref{aff104}}
\and G.~Polenta\orcid{0000-0003-4067-9196}\inst{\ref{aff105}}
\and M.~Poncet\inst{\ref{aff106}}
\and L.~A.~Popa\inst{\ref{aff107}}
\and L.~Pozzetti\orcid{0000-0001-7085-0412}\inst{\ref{aff17}}
\and F.~Raison\orcid{0000-0002-7819-6918}\inst{\ref{aff6}}
\and A.~Renzi\orcid{0000-0001-9856-1970}\inst{\ref{aff108},\ref{aff109},\ref{aff17}}
\and J.~Rhodes\orcid{0000-0002-4485-8549}\inst{\ref{aff79}}
\and G.~Riccio\inst{\ref{aff45}}
\and I.~Risso\orcid{0000-0003-2525-7761}\inst{\ref{aff56},\ref{aff57},\ref{aff50}}
\and F.~Rizzo\orcid{0000-0002-9407-585X}\inst{\ref{aff52}}
\and E.~Romelli\orcid{0000-0003-3069-9222}\inst{\ref{aff52}}
\and M.~Roncarelli\orcid{0000-0001-9587-7822}\inst{\ref{aff17}}
\and B.~Rusholme\orcid{0000-0001-7648-4142}\inst{\ref{aff25}}
\and R.~Saglia\orcid{0000-0003-0378-7032}\inst{\ref{aff5},\ref{aff6}}
\and Z.~Sakr\orcid{0000-0002-4823-3757}\inst{\ref{aff110},\ref{aff111},\ref{aff112}}
\and D.~Sapone\orcid{0000-0001-7089-4503}\inst{\ref{aff113}}
\and B.~Sartoris\orcid{0000-0003-1337-5269}\inst{\ref{aff6},\ref{aff52}}
\and P.~Schneider\orcid{0000-0001-8561-2679}\inst{\ref{aff91}}
\and T.~Schrabback\orcid{0000-0002-6987-7834}\inst{\ref{aff114}}
\and A.~Secroun\orcid{0000-0003-0505-3710}\inst{\ref{aff73}}
\and E.~Sihvola\orcid{0000-0003-1804-7715}\inst{\ref{aff115}}
\and P.~Simon\inst{\ref{aff91}}
\and C.~Sirignano\orcid{0000-0002-0995-7146}\inst{\ref{aff108},\ref{aff109}}
\and G.~Sirri\orcid{0000-0003-2626-2853}\inst{\ref{aff18}}
\and L.~Stanco\orcid{0000-0002-9706-5104}\inst{\ref{aff109}}
\and P.~Tallada-Cresp\'{i}\orcid{0000-0002-1336-8328}\inst{\ref{aff62},\ref{aff63}}
\and A.~N.~Taylor\inst{\ref{aff68}}
\and I.~Tereno\orcid{0000-0002-4537-6218}\inst{\ref{aff116},\ref{aff117}}
\and N.~Tessore\orcid{0000-0002-9696-7931}\inst{\ref{aff71}}
\and S.~Toft\orcid{0000-0003-3631-7176}\inst{\ref{aff118},\ref{aff119}}
\and R.~Toledo-Moreo\orcid{0000-0002-2997-4859}\inst{\ref{aff120},\ref{aff121}}
\and F.~Torradeflot\orcid{0000-0003-1160-1517}\inst{\ref{aff63},\ref{aff62}}
\and I.~Tutusaus\orcid{0000-0002-3199-0399}\inst{\ref{aff122},\ref{aff123},\ref{aff111}}
\and J.~Valiviita\orcid{0000-0001-6225-3693}\inst{\ref{aff9},\ref{aff87}}
\and T.~Vassallo\orcid{0000-0001-6512-6358}\inst{\ref{aff52},\ref{aff76}}
\and G.~Verdoes~Kleijn\orcid{0000-0001-5803-2580}\inst{\ref{aff30}}
\and A.~Veropalumbo\orcid{0000-0003-2387-1194}\inst{\ref{aff50},\ref{aff57},\ref{aff56}}
\and Y.~Wang\orcid{0000-0002-4749-2984}\inst{\ref{aff25}}
\and J.~Weller\orcid{0000-0002-8282-2010}\inst{\ref{aff5},\ref{aff6}}
\and A.~Zacchei\orcid{0000-0003-0396-1192}\inst{\ref{aff52},\ref{aff51}}
\and G.~Zamorani\orcid{0000-0002-2318-301X}\inst{\ref{aff17}}
\and F.~M.~Zerbi\orcid{0000-0002-9996-973X}\inst{\ref{aff50}}
\and E.~Zucca\orcid{0000-0002-5845-8132}\inst{\ref{aff17}}
\and M.~Ballardini\orcid{0000-0003-4481-3559}\inst{\ref{aff124},\ref{aff125},\ref{aff17}}
\and P.~Bergamini\orcid{0000-0003-1383-9414}\inst{\ref{aff17}}
\and C.~Burigana\orcid{0000-0002-3005-5796}\inst{\ref{aff24},\ref{aff75}}
\and R.~Cabanac\orcid{0000-0001-6679-2600}\inst{\ref{aff111}}
\and A.~Cappi\inst{\ref{aff126},\ref{aff17}}
\and F.~Caro\orcid{0009-0003-1053-0507}\inst{\ref{aff64}}
\and T.~Castro\orcid{0000-0002-6292-3228}\inst{\ref{aff127},\ref{aff52},\ref{aff51}}
\and D.~Di~Ferdinando\inst{\ref{aff18}}
\and J.~A.~Escartin~Vigo\inst{\ref{aff6}}
\and J.~Garc\'ia-Bellido\orcid{0000-0002-9370-8360}\inst{\ref{aff110}}
\and V.~Gautard\inst{\ref{aff128}}
\and C.~M.~Gutierrez\orcid{0000-0001-7854-783X}\inst{\ref{aff67},\ref{aff129}}
\and A.~Hall\orcid{0000-0002-3139-8651}\inst{\ref{aff68}}
\and A.~Loureiro\orcid{0000-0002-4371-0876}\inst{\ref{aff130},\ref{aff131}}
\and R.~Maoli\orcid{0000-0002-6065-3025}\inst{\ref{aff132},\ref{aff64}}
\and M.~P\"ontinen\orcid{0000-0001-5442-2530}\inst{\ref{aff9}}
\and E.~Sarpa\orcid{0000-0002-1256-655X}\inst{\ref{aff52}}
\and V.~Scottez\orcid{0009-0008-3864-940X}\inst{\ref{aff133},\ref{aff134}}
\and M.~Sereno\orcid{0000-0003-0302-0325}\inst{\ref{aff17},\ref{aff18}}
\and K.~Tanidis\orcid{0000-0001-9843-5130}\inst{\ref{aff135}}
\and M.~Tenti\orcid{0000-0002-4254-5901}\inst{\ref{aff18}}
\and M.~Tucci\inst{\ref{aff27}}
\and M.~Viel\orcid{0000-0002-2642-5707}\inst{\ref{aff51},\ref{aff52},\ref{aff54},\ref{aff53},\ref{aff136}}
\and M.~Wiesmann\orcid{0009-0000-8199-5860}\inst{\ref{aff78}}
\and J.~A.~Acevedo~Barroso\orcid{0000-0002-9654-1711}\inst{\ref{aff79}}
\and Y.~Akrami\orcid{0000-0002-2407-7956}\inst{\ref{aff110},\ref{aff137}}
\and I.~T.~Andika\orcid{0000-0001-6102-9526}\inst{\ref{aff5}}
\and G.~Angora\orcid{0000-0002-0316-6562}\inst{\ref{aff45},\ref{aff124}}
\and S.~Anselmi\orcid{0000-0002-3579-9583}\inst{\ref{aff109},\ref{aff108},\ref{aff138}}
\and M.~Archidiacono\orcid{0000-0003-4952-9012}\inst{\ref{aff139},\ref{aff140}}
\and G.~Aric\`o\orcid{0000-0002-2802-2928}\inst{\ref{aff18}}
\and F.~Atrio-Barandela\orcid{0000-0002-2130-2513}\inst{\ref{aff141}}
\and E.~Aubourg\orcid{0000-0002-5592-023X}\inst{\ref{aff94},\ref{aff142}}
\and M.~Baes\orcid{0000-0002-3930-2757}\inst{\ref{aff12}}
\and L.~Bazzanini\orcid{0000-0003-0727-0137}\inst{\ref{aff124},\ref{aff17}}
\and D.~Bertacca\orcid{0000-0002-2490-7139}\inst{\ref{aff108},\ref{aff77},\ref{aff109}}
\and M.~Bethermin\orcid{0000-0002-3915-2015}\inst{\ref{aff143}}
\and F.~Beutler\orcid{0000-0003-0467-5438}\inst{\ref{aff68}}
\and A.~Blanchard\orcid{0000-0001-8555-9003}\inst{\ref{aff111}}
\and L.~Blot\orcid{0000-0002-9622-7167}\inst{\ref{aff144},\ref{aff88}}
\and M.~Bonici\orcid{0000-0002-8430-126X}\inst{\ref{aff101},\ref{aff10}}
\and F.~Bouch\`e\orcid{0000-0002-4663-1786}\inst{\ref{aff145},\ref{aff65}}
\and M.~L.~Brown\orcid{0000-0002-0370-8077}\inst{\ref{aff146}}
\and S.~Bruton\orcid{0000-0002-6503-5218}\inst{\ref{aff147}}
\and A.~Calabro\orcid{0000-0003-2536-1614}\inst{\ref{aff64}}
\and B.~Camacho~Quevedo\orcid{0000-0002-8789-4232}\inst{\ref{aff51},\ref{aff54},\ref{aff52}}
\and C.~S.~Carvalho\inst{\ref{aff117}}
\and Y.~Charles\orcid{0009-0000-3636-834X}\inst{\ref{aff28}}
\and F.~Cogato\orcid{0000-0003-4632-6113}\inst{\ref{aff17}}
\and S.~Conseil\orcid{0000-0002-3657-4191}\inst{\ref{aff148},\ref{aff39}}
\and A.~R.~Cooray\orcid{0000-0002-3892-0190}\inst{\ref{aff149}}
\and P.~Corcho-Caballero\orcid{0000-0001-6327-7080}\inst{\ref{aff30}}
\and M.~Correnti\orcid{0000-0001-6464-3257}\inst{\ref{aff64},\ref{aff105}}
\and B.~Csizi\orcid{0000-0003-3227-6581}\inst{\ref{aff114}}
\and O.~Cucciati\orcid{0000-0002-9336-7551}\inst{\ref{aff17}}
\and T.~de~Boer\orcid{0000-0001-5486-2747}\inst{\ref{aff20}}
\and G.~Desprez\orcid{0000-0001-8325-1742}\inst{\ref{aff30}}
\and A.~D\'iaz-S\'anchez\orcid{0000-0003-0748-4768}\inst{\ref{aff150}}
\and S.~Di~Domizio\orcid{0000-0003-2863-5895}\inst{\ref{aff56},\ref{aff57}}
\and J.~M.~Diego\orcid{0000-0001-9065-3926}\inst{\ref{aff33}}
\and V.~Duret\orcid{0009-0009-0383-4960}\inst{\ref{aff73}}
\and M.~Y.~Elkhashab\orcid{0000-0001-9306-2603}\inst{\ref{aff52},\ref{aff53},\ref{aff97},\ref{aff51}}
\and Y.~Fang\orcid{0000-0002-0334-6950}\inst{\ref{aff5}}
\and A.~Farina\orcid{0009-0000-3420-929X}\inst{\ref{aff50},\ref{aff57}}
\and A.~Finoguenov\orcid{0000-0002-4606-5403}\inst{\ref{aff9}}
\and A.~Franco\orcid{0000-0002-4761-366X}\inst{\ref{aff151},\ref{aff152},\ref{aff153}}
\and A.~Fumagalli\orcid{0009-0004-0300-2535}\inst{\ref{aff52}}
\and Y.~Fu\orcid{0000-0002-0759-0504}\inst{\ref{aff154},\ref{aff30}}
\and K.~Ganga\orcid{0000-0001-8159-8208}\inst{\ref{aff94}}
\and E.~Gaztanaga\orcid{0000-0001-9632-0815}\inst{\ref{aff122},\ref{aff123},\ref{aff1}}
\and Z.~Ghaffari\orcid{0000-0002-6467-8078}\inst{\ref{aff52},\ref{aff51}}
\and F.~Giacomini\orcid{0000-0002-3129-2814}\inst{\ref{aff18}}
\and F.~Gianotti\orcid{0000-0003-4666-119X}\inst{\ref{aff17}}
\and G.~Gozaliasl\orcid{0000-0002-0236-919X}\inst{\ref{aff155},\ref{aff9}}
\and A.~Gruppuso\orcid{0000-0001-9272-5292}\inst{\ref{aff17},\ref{aff18}}
\and M.~Guidi\orcid{0000-0001-9408-1101}\inst{\ref{aff38},\ref{aff17}}
\and C.~Hern\'andez-Monteagudo\orcid{0000-0001-5471-9166}\inst{\ref{aff129},\ref{aff67}}
\and T.~Heuberger\orcid{0009-0009-3474-8467}\inst{\ref{aff19}}
\and H.~Hildebrandt\orcid{0000-0002-9814-3338}\inst{\ref{aff156}}
\and J.~Hjorth\orcid{0000-0002-4571-2306}\inst{\ref{aff157}}
\and M.~Jauzac\orcid{0000-0003-1974-8732}\inst{\ref{aff158},\ref{aff93},\ref{aff159},\ref{aff160},\ref{aff111}}
\and J.~J.~E.~Kajava\orcid{0000-0002-3010-8333}\inst{\ref{aff161},\ref{aff8},\ref{aff162}}
\and Y.~Kang\orcid{0009-0000-8588-7250}\inst{\ref{aff114}}
\and V.~Kansal\orcid{0000-0002-4008-6078}\inst{\ref{aff163},\ref{aff164}}
\and D.~Karagiannis\orcid{0000-0002-4927-0816}\inst{\ref{aff124},\ref{aff165}}
\and J.~Kim\orcid{0000-0003-2776-2761}\inst{\ref{aff23}}
\and C.~C.~Kirkpatrick\inst{\ref{aff115}}
\and A.~Kov\'acs\orcid{0000-0002-5825-579X}\inst{\ref{aff166},\ref{aff167}}
\and I.~Kova{\v{c}}i{\'{c}}\orcid{0000-0001-6751-3263}\inst{\ref{aff12}}
\and K.~Koyama\orcid{0000-0001-6727-6915}\inst{\ref{aff1}}
\and S.~Kruk\orcid{0000-0001-8010-8879}\inst{\ref{aff48}}
\and M.~C.~Lam\orcid{0000-0002-9347-2298}\inst{\ref{aff68}}
\and M.~Lattanzi\orcid{0000-0003-1059-2532}\inst{\ref{aff125}}
\and S.~Lee\orcid{0000-0002-8289-740X}\inst{\ref{aff168},\ref{aff79}}
\and G.~Leroy\orcid{0009-0004-2523-4425}\inst{\ref{aff158},\ref{aff93}}
\and G.~F.~Lesci\orcid{0000-0002-4607-2830}\inst{\ref{aff16},\ref{aff17}}
\and J.~Lesgourgues\orcid{0000-0001-7627-353X}\inst{\ref{aff169}}
\and L.~Linke\orcid{0000-0002-2622-8113}\inst{\ref{aff114}}
\and G.~Maggio\orcid{0000-0003-4020-4836}\inst{\ref{aff52}}
\and A.~Manj\'on-Garc\'ia\orcid{0000-0002-7413-8825}\inst{\ref{aff150}}
\and F.~Mannucci\orcid{0000-0002-4803-2381}\inst{\ref{aff170}}
\and F.~R.~Marleau\orcid{0000-0002-1442-2947}\inst{\ref{aff114}}
\and C.~J.~A.~P.~Martins\orcid{0000-0002-4886-9261}\inst{\ref{aff171},\ref{aff172}}
\and M.~Miluzio\inst{\ref{aff48},\ref{aff173}}
\and G.~Morgante\inst{\ref{aff17}}
\and K.~Naidoo\orcid{0000-0002-9182-1802}\inst{\ref{aff1},\ref{aff85}}
\and A.~Navarro-Alsina\orcid{0000-0002-3173-2592}\inst{\ref{aff91}}
\and S.~Nesseris\orcid{0000-0002-0567-0324}\inst{\ref{aff110}}
\and L.~Nicastro\orcid{0000-0001-8534-6788}\inst{\ref{aff17}}
\and F.~Oppizzi\orcid{0000-0003-3904-8370}\inst{\ref{aff57},\ref{aff109},\ref{aff108}}
\and F.~Pace\orcid{0000-0001-8039-0480}\inst{\ref{aff59},\ref{aff60},\ref{aff61},\ref{aff174}}
\and D.~Paoletti\orcid{0000-0003-4761-6147}\inst{\ref{aff17},\ref{aff75}}
\and F.~Passalacqua\orcid{0000-0002-8606-4093}\inst{\ref{aff109}}
\and K.~Paterson\orcid{0000-0001-8340-3486}\inst{\ref{aff85}}
\and L.~Patrizii\inst{\ref{aff18}}
\and C.~Pattison\orcid{0000-0003-3272-2617}\inst{\ref{aff1}}
\and A.~Pisani\orcid{0000-0002-6146-4437}\inst{\ref{aff73}}
\and L.~Pizzuti\orcid{0000-0001-5654-7580}\inst{\ref{aff175}}
\and J.~J.~Popp\orcid{0000-0002-3724-1727}\inst{\ref{aff176}}
\and D.~Potter\orcid{0000-0002-0757-5195}\inst{\ref{aff177}}
\and G.~W.~Pratt\inst{\ref{aff104}}
\and S.~Quai\orcid{0000-0002-0449-8163}\inst{\ref{aff16},\ref{aff17}}
\and M.~Radovich\orcid{0000-0002-3585-866X}\inst{\ref{aff77}}
\and G.~Rodighiero\orcid{0000-0002-9415-2296}\inst{\ref{aff108},\ref{aff77}}
\and W.~Roster\orcid{0000-0002-9149-6528}\inst{\ref{aff6}}
\and M.~Sahl\'en\orcid{0000-0003-0973-4804}\inst{\ref{aff178}}
\and D.~B.~Sanders\orcid{0000-0002-1233-9998}\inst{\ref{aff20}}
\and A.~Schneider\orcid{0000-0001-7055-8104}\inst{\ref{aff177}}
\and D.~Sciotti\orcid{0009-0008-4519-2620}\inst{\ref{aff64},\ref{aff92}}
\and E.~Sellentin\orcid{0009-0002-2655-3458}\inst{\ref{aff179},\ref{aff154}}
\and S.~Serjeant\orcid{0000-0002-0517-7943}\inst{\ref{aff176}}
\and L.~C.~Smith\orcid{0000-0002-3259-2771}\inst{\ref{aff31}}
\and J.~G.~Sorce\orcid{0000-0002-2307-2432}\inst{\ref{aff180},\ref{aff72}}
\and I.~Szapudi\orcid{0000-0003-2274-0301}\inst{\ref{aff20}}
\and M.~Talia\orcid{0000-0003-4352-2063}\inst{\ref{aff16},\ref{aff17}}
\and C.~Tao\orcid{0000-0001-7961-8177}\inst{\ref{aff73}}
\and F.~Tarsitano\orcid{0000-0002-5919-0238}\inst{\ref{aff181},\ref{aff182}}
\and G.~Testera\orcid{0000-0003-2970-766X}\inst{\ref{aff57}}
\and R.~Teyssier\orcid{0000-0001-7689-0933}\inst{\ref{aff183}}
\and S.~Tosi\orcid{0000-0002-7275-9193}\inst{\ref{aff56},\ref{aff50},\ref{aff57}}
\and A.~Troja\orcid{0000-0003-0239-4595}\inst{\ref{aff52}}
\and C.~Uhlemann\orcid{0000-0001-7831-1579}\inst{\ref{aff184},\ref{aff4}}
\and C.~Valieri\inst{\ref{aff18}}
\and A.~Venhola\orcid{0000-0001-6071-4564}\inst{\ref{aff185}}
\and D.~Vergani\orcid{0000-0003-0898-2216}\inst{\ref{aff17}}
\and G.~Vernardos\orcid{0000-0001-8554-7248}\inst{\ref{aff22},\ref{aff186}}
\and P.~Vielzeuf\orcid{0000-0003-2035-9339}\inst{\ref{aff73}}
\and S.~Vinciguerra\orcid{0009-0005-4018-3184}\inst{\ref{aff28}}
\and M.~von~Wietersheim-Kramsta\orcid{0000-0003-4986-5091}\inst{\ref{aff93},\ref{aff158}}
\and N.~A.~Walton\orcid{0000-0003-3983-8778}\inst{\ref{aff31}}
\and L.~Wang\orcid{0000-0002-6736-9158}\inst{\ref{aff187},\ref{aff30}}
\and J.~Wilde\orcid{0000-0002-4460-7379}\inst{\ref{aff13}}
\and A.~H.~Wright\orcid{0000-0001-7363-7932}\inst{\ref{aff156}}
\and D.~Scott\orcid{0000-0002-6878-9840}\inst{\ref{aff43}}}
										   
\institute{Institute of Cosmology and Gravitation, University of Portsmouth, Portsmouth PO1 3FX, UK\label{aff1}
\and
Fontys University of Applied Sciences, School of ICT, Rachelsmolen 1, 5612 MA Eindhoven, The Netherlands\label{aff2}
\and
University of Applied Sciences and Arts of Northwestern Switzerland, School of Engineering, 5210 Windisch, Switzerland\label{aff3}
\and
School of Mathematics, Statistics and Physics, Newcastle University, Herschel Building, Newcastle-upon-Tyne, NE1 7RU, UK\label{aff4}
\and
Universit\"ats-Sternwarte M\"unchen, Fakult\"at f\"ur Physik, Ludwig-Maximilians-Universit\"at M\"unchen, Scheinerstr.~1, 81679 M\"unchen, Germany\label{aff5}
\and
Max Planck Institute for Extraterrestrial Physics, Giessenbachstr. 1, 85748 Garching, Germany\label{aff6}
\and
Max-Planck-Institut f\"ur Astrophysik, Karl-Schwarzschild-Str.~1, 85748 Garching, Germany\label{aff7}
\and
Finnish Centre for Astronomy with ESO (FINCA), Quantum, Vesilinnantie 5, University of Turku, 20014 Turku, Finland\label{aff8}
\and
Department of Physics, P.O. Box 64, University of Helsinki, 00014 Helsinki, Finland\label{aff9}
\and
INAF-IASF Milano, Via Alfonso Corti 12, 20133 Milano, Italy\label{aff10}
\and
STAR Institute, University of Li{\`e}ge, Quartier Agora, All\'ee du six Ao\^ut 19c, 4000 Li\`ege, Belgium\label{aff11}
\and
Universiteit Gent, Department of Physics and Astronomy, Proeftuinstraat 86 N3, 9000 Ghent, Belgium
\label{aff12}
\and
Institut de Ci\`{e}ncies del Cosmos (ICCUB), Universitat de Barcelona (IEEC-UB), Mart\'{i} i Franqu\`{e}s 1, 08028 Barcelona, Spain\label{aff13}
\and
Instituci\'o Catalana de Recerca i Estudis Avan\c{c}ats (ICREA), Passeig de Llu\'{\i}s Companys 23, 08010 Barcelona, Spain\label{aff14}
\and
Institut de Ciencies de l'Espai (IEEC-CSIC), Campus UAB, Carrer de Can Magrans, s/n Cerdanyola del Vall\'es, 08193 Barcelona, Spain\label{aff15}
\and
Dipartimento di Fisica e Astronomia "Augusto Righi" - Alma Mater Studiorum Universit\`a di Bologna, via Piero Gobetti 93/2, 40129 Bologna, Italy\label{aff16}
\and
INAF-Osservatorio di Astrofisica e Scienza dello Spazio di Bologna, Via Piero Gobetti 93/3, 40129 Bologna, Italy\label{aff17}
\and
INFN-Sezione di Bologna, Viale Berti Pichat 6/2, 40127 Bologna, Italy\label{aff18}
\and
University of Applied Sciences and Arts of Northwestern Switzerland, School of Computer Science, 5210 Windisch, Switzerland\label{aff19}
\and
Institute for Astronomy, University of Hawaii, 2680 Woodlawn Drive, Honolulu, HI 96822, USA\label{aff20}
\and
CUNY Graduate Center, City University of New York, 365 5th Ave., New York, NY 10016, USA\label{aff21}
\and
Department of Physics and Astronomy, Lehman College of the CUNY, Bronx, NY 10468, USA\label{aff22}
\and
Department of Physics, University of Oxford, Keble Road, Oxford OX1 3RH, UK\label{aff23}
\and
INAF, Istituto di Radioastronomia, Via Piero Gobetti 101, 40129 Bologna, Italy\label{aff24}
\and
Caltech/IPAC, 1200 E. California Blvd., Pasadena, CA 91125, USA\label{aff25}
\and
Laboratoire univers et particules de Montpellier, Universit\'e de Montpellier, CNRS, 34090 Montpellier, France\label{aff26}
\and
Department of Astronomy, University of Geneva, ch. d'Ecogia 16, 1290 Versoix, Switzerland\label{aff27}
\and
Aix-Marseille Universit\'e, CNRS, CNES, LAM, Marseille, France\label{aff28}
\and
Institut d'Astrophysique de Paris, UMR 7095, CNRS, and Sorbonne Universit\'e, 98 bis boulevard Arago, 75014 Paris, France\label{aff29}
\and
Kapteyn Astronomical Institute, University of Groningen, PO Box 800, 9700 AV Groningen, The Netherlands\label{aff30}
\and
Institute of Astronomy, University of Cambridge, Madingley Road, Cambridge CB3 0HA, UK\label{aff31}
\and
Kavli Institute for Cosmology Cambridge, Madingley Road, Cambridge, CB3 0HA, UK\label{aff32}
\and
Instituto de F\'isica de Cantabria, Edificio Juan Jord\'a, Avenida de los Castros, 39005 Santander, Spain\label{aff33}
\and
European Southern Observatory, Karl-Schwarzschild-Str.~2, 85748 Garching, Germany\label{aff34}
\and
Departamento de F\'isica, Universidad de Santiago de Chile, Av. Victor Jara 3493, Santiago, Chile\label{aff35}
\and
Center for Interdisciplinary Research in Astrophysics Space Science (CIRAS), Universidad de Santiago, Chile\label{aff36}
\and
Universit\'e de Gen\`eve, D\'epartement de Physique Th\'eorique and Centre for Astroparticle Physics, 24 quai Ernest-Ansermet, CH-1211 Gen\`eve 4, Switzerland\label{aff37}
\and
Dipartimento di Fisica e Astronomia, Universit\`a di Bologna, Via Gobetti 93/2, 40129 Bologna, Italy\label{aff38}
\and
Universit\'e Lyon 1, CNRS/IN2P3, IP2I Lyon, Villeurbanne, France\label{aff39}
\and
Department of Astronomy, School of Physics and Astronomy, Shanghai Jiao Tong University, Shanghai 200240, China\label{aff40}
\and
State Key Laboratory of Dark Matter Physics, School of Physics and Astronomy, Shanghai Jiao Tong University, Shanghai 200240, China\label{aff41}
\and
Shanghai Key Laboratory for Particle Physics and Cosmology, Shanghai 200240, China\label{aff42}
\and
Department of Physics and Astronomy, University of British Columbia, Vancouver, BC V6T 1Z1, Canada\label{aff43}
\and
National Astronomical Observatory of Japan, 2-21-1 Osawa, Mitaka, Tokyo 181-8588, Japan\label{aff44}
\and
INAF-Osservatorio Astronomico di Capodimonte, Via Moiariello 16, 80131 Napoli, Italy\label{aff45}
\and
Technical University of Munich, TUM School of Natural Sciences, Physics Department, James-Franck-Str.~1, 85748 Garching, Germany\label{aff46}
\and
Shanghai Astronomical Observatory (SHAO), Nandan Road 80, Shanghai 200030, China\label{aff47}
\and
ESAC/ESA, Camino Bajo del Castillo, s/n., Urb. Villafranca del Castillo, 28692 Villanueva de la Ca\~nada, Madrid, Spain\label{aff48}
\and
School of Mathematics and Physics, University of Surrey, Guildford, Surrey, GU2 7XH, UK\label{aff49}
\and
INAF-Osservatorio Astronomico di Brera, Via Brera 28, 20122 Milano, Italy\label{aff50}
\and
IFPU, Institute for Fundamental Physics of the Universe, via Beirut 2, 34151 Trieste, Italy\label{aff51}
\and
INAF-Osservatorio Astronomico di Trieste, Via G. B. Tiepolo 11, 34143 Trieste, Italy\label{aff52}
\and
INFN, Sezione di Trieste, Via Valerio 2, 34127 Trieste TS, Italy\label{aff53}
\and
SISSA, International School for Advanced Studies, Via Bonomea 265, 34136 Trieste TS, Italy\label{aff54}
\and
Institut de Physique Th\'eorique, CEA, CNRS, Universit\'e Paris-Saclay 91191 Gif-sur-Yvette Cedex, France\label{aff55}
\and
Dipartimento di Fisica, Universit\`a di Genova, Via Dodecaneso 33, 16146, Genova, Italy\label{aff56}
\and
INFN-Sezione di Genova, Via Dodecaneso 33, 16146, Genova, Italy\label{aff57}
\and
Department of Physics "E. Pancini", University Federico II, Via Cinthia 6, 80126, Napoli, Italy\label{aff58}
\and
Dipartimento di Fisica, Universit\`a degli Studi di Torino, Via P. Giuria 1, 10125 Torino, Italy\label{aff59}
\and
INFN-Sezione di Torino, Via P. Giuria 1, 10125 Torino, Italy\label{aff60}
\and
INAF-Osservatorio Astrofisico di Torino, Via Osservatorio 20, 10025 Pino Torinese (TO), Italy\label{aff61}
\and
Centro de Investigaciones Energ\'eticas, Medioambientales y Tecnol\'ogicas (CIEMAT), Avenida Complutense 40, 28040 Madrid, Spain\label{aff62}
\and
Port d'Informaci\'{o} Cient\'{i}fica, Campus UAB, C. Albareda s/n, 08193 Bellaterra (Barcelona), Spain\label{aff63}
\and
INAF-Osservatorio Astronomico di Roma, Via Frascati 33, 00078 Monteporzio Catone, Italy\label{aff64}
\and
INFN -- Sezione di Napoli, Via Cinthia 6, 80126, Napoli, Italy\label{aff65}
\and
Dipartimento di Fisica e Astronomia "Augusto Righi" - Alma Mater Studiorum Universit\`a di Bologna, Viale Berti Pichat 6/2, 40127 Bologna, Italy\label{aff66}
\and
Instituto de Astrof\'{\i}sica de Canarias, E-38205 La Laguna, Tenerife, Spain\label{aff67}
\and
Institute for Astronomy, University of Edinburgh, Royal Observatory, Blackford Hill, Edinburgh EH9 3HJ, UK\label{aff68}
\and
European Space Agency/ESRIN, Largo Galileo Galilei 1, 00044 Frascati, Roma, Italy\label{aff69}
\and
Universit\'e Lyon 1, CNRS/IN2P3, IUF, IP2I Lyon, Villeurbanne, France \label{aff70}
\and
Mullard Space Science Laboratory, University College London, Holmbury St Mary, Dorking, Surrey RH5 6NT, UK\label{aff71}
\and
Universit\'e Paris-Saclay, CNRS, Institut d'astrophysique spatiale, 91405, Orsay, France\label{aff72}
\and
Aix-Marseille Universit\'e, CNRS/IN2P3, CPPM, Marseille, France\label{aff73}
\and
INAF-Istituto di Astrofisica e Planetologia Spaziali, via del Fosso del Cavaliere, 100, 00100 Roma, Italy\label{aff74}
\and
INFN-Bologna, Via Irnerio 46, 40126 Bologna, Italy\label{aff75}
\and
University Observatory, LMU Faculty of Physics, Scheinerstr.~1, 81679 Munich, Germany\label{aff76}
\and
INAF-Osservatorio Astronomico di Padova, Via dell'Osservatorio 5, 35122 Padova, Italy\label{aff77}
\and
Institute of Theoretical Astrophysics, University of Oslo, P.O. Box 1029 Blindern, 0315 Oslo, Norway\label{aff78}
\and
Jet Propulsion Laboratory, California Institute of Technology, 4800 Oak Grove Drive, Pasadena, CA, 91109, USA\label{aff79}
\and
Felix Hormuth Engineering, Goethestr. 17, 69181 Leimen, Germany\label{aff80}
\and
Technical University of Denmark, Elektrovej 327, 2800 Kgs. Lyngby, Denmark\label{aff81}
\and
Cosmic Dawn Center (DAWN), Denmark\label{aff82}
\and
Universit\'e PSL, Observatoire de Paris, Sorbonne Universit\'e, CNRS, LERMA, 75014, Paris, France\label{aff83}
\and
Universit\'e Paris-Cit\'e, 5 Rue Thomas Mann, 75013, Paris, France\label{aff84}
\and
Max-Planck-Institut f\"ur Astronomie, K\"onigstuhl 17, 69117 Heidelberg, Germany\label{aff85}
\and
NASA Goddard Space Flight Center, Greenbelt, MD 20771, USA\label{aff86}
\and
Helsinki Institute of Physics, Gustaf H{\"a}llstr{\"o}min katu 2, University of Helsinki, 00014 Helsinki, Finland\label{aff87}
\and
Laboratoire d'etude de l'Univers et des phenomenes eXtremes, Observatoire de Paris, Universit\'e PSL, Sorbonne Universit\'e, CNRS, 92190 Meudon, France\label{aff88}
\and
SKAO, Jodrell Bank, Lower Withington, Macclesfield SK11 9FT, UK\label{aff89}
\and
Centre de Calcul de l'IN2P3/CNRS, 21 avenue Pierre de Coubertin 69627 Villeurbanne Cedex, France\label{aff90}
\and
Universit\"at Bonn, Argelander-Institut f\"ur Astronomie, Auf dem H\"ugel 71, 53121 Bonn, Germany\label{aff91}
\and
INFN-Sezione di Roma, Piazzale Aldo Moro, 2 - c/o Dipartimento di Fisica, Edificio G. Marconi, 00185 Roma, Italy\label{aff92}
\and
Department of Physics, Institute for Computational Cosmology, Durham University, South Road, Durham, DH1 3LE, UK\label{aff93}
\and
Universit\'e Paris Cit\'e, CNRS, Astroparticule et Cosmologie, 75013 Paris, France\label{aff94}
\and
CNRS-UCB International Research Laboratory, Centre Pierre Bin\'etruy, IRL2007, CPB-IN2P3, Berkeley, USA\label{aff95}
\and
Institute of Physics, Laboratory of Astrophysics, Ecole Polytechnique F\'ed\'erale de Lausanne (EPFL), Observatoire de Sauverny, 1290 Versoix, Switzerland\label{aff96}
\and
Dipartimento di Fisica - Sezione di Astronomia, Universit\`a di Trieste, Via Tiepolo 11, 34131 Trieste, Italy\label{aff97}
\and
Telespazio UK S.L. for European Space Agency (ESA), Camino bajo del Castillo, s/n, Urbanizacion Villafranca del Castillo, Villanueva de la Ca\~nada, 28692 Madrid, Spain\label{aff98}
\and
Institut de F\'{i}sica d'Altes Energies (IFAE), The Barcelona Institute of Science and Technology, Campus UAB, 08193 Bellaterra (Barcelona), Spain\label{aff99}
\and
European Space Agency/ESTEC, Keplerlaan 1, 2201 AZ Noordwijk, The Netherlands\label{aff100}
\and
Waterloo Centre for Astrophysics, University of Waterloo, Waterloo, Ontario N2L 3G1, Canada\label{aff101}
\and
Department of Physics and Astronomy, University of Waterloo, Waterloo, Ontario N2L 3G1, Canada\label{aff102}
\and
Perimeter Institute for Theoretical Physics, Waterloo, Ontario N2L 2Y5, Canada\label{aff103}
\and
Universit\'e Paris-Saclay, Universit\'e Paris Cit\'e, CEA, CNRS, AIM, 91191, Gif-sur-Yvette, France\label{aff104}
\and
Space Science Data Center, Italian Space Agency, via del Politecnico snc, 00133 Roma, Italy\label{aff105}
\and
Centre National d'Etudes Spatiales -- Centre spatial de Toulouse, 18 avenue Edouard Belin, 31401 Toulouse Cedex 9, France\label{aff106}
\and
Institute of Space Science, Str. Atomistilor, nr. 409 M\u{a}gurele, Ilfov, 077125, Romania\label{aff107}
\and
Dipartimento di Fisica e Astronomia "G. Galilei", Universit\`a di Padova, Via Marzolo 8, 35131 Padova, Italy\label{aff108}
\and
INFN-Padova, Via Marzolo 8, 35131 Padova, Italy\label{aff109}
\and
Instituto de F\'isica Te\'orica UAM-CSIC, Campus de Cantoblanco, 28049 Madrid, Spain\label{aff110}
\and
Institut de Recherche en Astrophysique et Plan\'etologie (IRAP), Universit\'e de Toulouse, CNRS, UPS, CNES, 14 Av. Edouard Belin, 31400 Toulouse, France\label{aff111}
\and
Universit\'e St Joseph; Faculty of Sciences, Beirut, Lebanon\label{aff112}
\and
Departamento de F\'isica, FCFM, Universidad de Chile, Blanco Encalada 2008, Santiago, Chile\label{aff113}
\and
Universit\"at Innsbruck, Institut f\"ur Astro- und Teilchenphysik, Technikerstr. 25/8, 6020 Innsbruck, Austria\label{aff114}
\and
Department of Physics and Helsinki Institute of Physics, Gustaf H\"allstr\"omin katu 2, University of Helsinki, 00014 Helsinki, Finland\label{aff115}
\and
Departamento de F\'isica, Faculdade de Ci\^encias, Universidade de Lisboa, Edif\'icio C8, Campo Grande, PT1749-016 Lisboa, Portugal\label{aff116}
\and
Instituto de Astrof\'isica e Ci\^encias do Espa\c{c}o, Faculdade de Ci\^encias, Universidade de Lisboa, Tapada da Ajuda, 1349-018 Lisboa, Portugal\label{aff117}
\and
Cosmic Dawn Center (DAWN)\label{aff118}
\and
Niels Bohr Institute, University of Copenhagen, Jagtvej 128, 2200 Copenhagen, Denmark\label{aff119}
\and
Universidad Polit\'ecnica de Cartagena, Departamento de Electr\'onica y Tecnolog\'ia de Computadoras,  Plaza del Hospital 1, 30202 Cartagena, Spain\label{aff120}
\and
European University of Technology EUt+, European Union\label{aff121}
\and
Institute of Space Sciences (ICE, CSIC), Campus UAB, Carrer de Can Magrans, s/n, 08193 Barcelona, Spain\label{aff122}
\and
Institut d'Estudis Espacials de Catalunya (IEEC),  Edifici RDIT, Campus UPC, 08860 Castelldefels, Barcelona, Spain\label{aff123}
\and
Dipartimento di Fisica e Scienze della Terra, Universit\`a degli Studi di Ferrara, Via Giuseppe Saragat 1, 44122 Ferrara, Italy\label{aff124}
\and
Istituto Nazionale di Fisica Nucleare, Sezione di Ferrara, Via Giuseppe Saragat 1, 44122 Ferrara, Italy\label{aff125}
\and
Universit\'e C\^{o}te d'Azur, Observatoire de la C\^{o}te d'Azur, CNRS, Laboratoire Lagrange, Bd de l'Observatoire, CS 34229, 06304 Nice cedex 4, France\label{aff126}
\and
Department of Mathematical Physics, Institute of Physics, University of S\~ao Paulo, R. do Mat\~ao 1371, 05508-090, S\~ao Paulo, SP, Brazil\label{aff127}
\and
CEA Saclay, DFR/IRFU, Service d'Astrophysique, Bat. 709, 91191 Gif-sur-Yvette, France\label{aff128}
\and
Universidad de La Laguna, Dpto. Astrof\'\i sica, E-38206 La Laguna, Tenerife, Spain\label{aff129}
\and
Oskar Klein Centre for Cosmoparticle Physics, Department of Physics, Stockholm University, Stockholm, SE-106 91, Sweden\label{aff130}
\and
Astrophysics Group, Blackett Laboratory, Imperial College London, London SW7 2AZ, UK\label{aff131}
\and
Dipartimento di Fisica, Sapienza Universit\`a di Roma, Piazzale Aldo Moro 2, 00185 Roma, Italy\label{aff132}
\and
Institut d'Astrophysique de Paris, 98bis Boulevard Arago, 75014, Paris, France\label{aff133}
\and
ICL, Junia, Universit\'e Catholique de Lille, LITL, 59000 Lille, France\label{aff134}
\and
Center for Astrophysics and Cosmology, University of Nova Gorica, Nova Gorica, Slovenia\label{aff135}
\and
ICSC - Centro Nazionale di Ricerca in High Performance Computing, Big Data e Quantum Computing, Via Magnanelli 2, Bologna, Italy\label{aff136}
\and
CERCA/ISO, Department of Physics, Case Western Reserve University, 10900 Euclid Avenue, Cleveland, OH 44106, USA\label{aff137}
\and
Laboratoire Univers et Th\'eorie, Observatoire de Paris, Universit\'e PSL, Universit\'e Paris Cit\'e, CNRS, 92190 Meudon, France\label{aff138}
\and
Dipartimento di Fisica "Aldo Pontremoli", Universit\`a degli Studi di Milano, Via Celoria 16, 20133 Milano, Italy\label{aff139}
\and
INFN-Sezione di Milano, Via Celoria 16, 20133 Milano, Italy\label{aff140}
\and
Departamento de F{\'\i}sica Fundamental. Universidad de Salamanca. Plaza de la Merced s/n. 37008 Salamanca, Spain\label{aff141}
\and
IRFU, CEA, Universit\'e Paris-Saclay 91191 Gif-sur-Yvette Cedex, France\label{aff142}
\and
Universit\'e de Strasbourg, CNRS, Observatoire astronomique de Strasbourg, UMR 7550, 67000 Strasbourg, France\label{aff143}
\and
Center for Data-Driven Discovery, Kavli IPMU (WPI), UTIAS, The University of Tokyo, Kashiwa, Chiba 277-8583, Japan\label{aff144}
\and
Scuola Superiore Meridionale, Via Mezzocannone 4, 80138, Napoli, Italy\label{aff145}
\and
Jodrell Bank Centre for Astrophysics, Department of Physics and Astronomy, University of Manchester, Oxford Road, Manchester M13 9PL, UK\label{aff146}
\and
California Institute of Technology, 1200 E California Blvd, Pasadena, CA 91125, USA\label{aff147}
\and
Centre de Recherche Astrophysique de Lyon, UMR5574, CNRS, Universit\'e Claude Bernard Lyon 1, ENS de Lyon, 69230, Saint-Genis-Laval, France\label{aff148}
\and
Department of Physics \& Astronomy, University of California Irvine, Irvine CA 92697, USA\label{aff149}
\and
Departamento F\'isica Aplicada, Universidad Polit\'ecnica de Cartagena, Campus Muralla del Mar, 30202 Cartagena, Murcia, Spain\label{aff150}
\and
Department of Mathematics and Physics E. De Giorgi, University of Salento, Via per Arnesano, CP-I93, 73100, Lecce, Italy\label{aff151}
\and
INFN, Sezione di Lecce, Via per Arnesano, CP-193, 73100, Lecce, Italy\label{aff152}
\and
INAF-Sezione di Lecce, c/o Dipartimento Matematica e Fisica, Via per Arnesano, 73100, Lecce, Italy\label{aff153}
\and
Leiden Observatory, Leiden University, Einsteinweg 55, 2333 CC Leiden, The Netherlands\label{aff154}
\and
Department of Computer Science, Aalto University, PO Box 15400, Espoo, FI-00 076, Finland\label{aff155}
\and
Ruhr University Bochum, Faculty of Physics and Astronomy, Astronomical Institute (AIRUB), German Centre for Cosmological Lensing (GCCL), 44780 Bochum, Germany\label{aff156}
\and
DARK, Niels Bohr Institute, University of Copenhagen, Jagtvej 155, 2200 Copenhagen, Denmark\label{aff157}
\and
Department of Physics, Centre for Extragalactic Astronomy, Durham University, South Road, Durham, DH1 3LE, UK\label{aff158}
\and
Astrophysics Research Centre, University of KwaZulu-Natal, Westville Campus, Durban 4041, South Africa\label{aff159}
\and
School of Mathematics, Statistics \& Computer Science, University of KwaZulu-Natal, Westville Campus, Durban 4041, South Africa\label{aff160}
\and
Department of Physics and Astronomy, Vesilinnantie 5, University of Turku, 20014 Turku, Finland\label{aff161}
\and
Serco for European Space Agency (ESA), Camino bajo del Castillo, s/n, Urbanizacion Villafranca del Castillo, Villanueva de la Ca\~nada, 28692 Madrid, Spain\label{aff162}
\and
ARC Centre of Excellence for Dark Matter Particle Physics, Melbourne, Australia\label{aff163}
\and
Centre for Astrophysics \& Supercomputing, Swinburne University of Technology,  Hawthorn, Victoria 3122, Australia\label{aff164}
\and
Department of Physics and Astronomy, University of the Western Cape, Bellville, Cape Town, 7535, South Africa\label{aff165}
\and
MTA-CSFK Lend\"ulet Large-Scale Structure Research Group, Konkoly-Thege Mikl\'os \'ut 15-17, H-1121 Budapest, Hungary\label{aff166}
\and
Konkoly Observatory, HUN-REN CSFK, MTA Centre of Excellence, Budapest, Konkoly Thege Mikl\'os {\'u}t 15-17. H-1121, Hungary\label{aff167}
\and
Ohio University, Physics \& Astronomy Department,1 Ohio University, Athens, OH 45701, USA\label{aff168}
\and
Institute for Theoretical Particle Physics and Cosmology (TTK), RWTH Aachen University, 52056 Aachen, Germany\label{aff169}
\and
INAF-Osservatorio Astrofisico di Arcetri, Largo E. Fermi 5, 50125, Firenze, Italy\label{aff170}
\and
Centro de Astrof\'{\i}sica da Universidade do Porto, Rua das Estrelas, 4150-762 Porto, Portugal\label{aff171}
\and
Instituto de Astrof\'isica e Ci\^encias do Espa\c{c}o, Universidade do Porto, CAUP, Rua das Estrelas, PT4150-762 Porto, Portugal\label{aff172}
\and
HE Space for European Space Agency (ESA), Camino bajo del Castillo, s/n, Urbanizacion Villafranca del Castillo, Villanueva de la Ca\~nada, 28692 Madrid, Spain\label{aff173}
\and
Instituto de Astrof\'isica e Ci\^encias do Espa\c{c}o, Faculdade de Ci\^encias, Universidade de Lisboa, Campo Grande, 1749-016 Lisboa, Portugal\label{aff174}
\and
Dipartimento di Fisica ``G. Occhialini", Universit\`a degli Studi di Milano Bicocca, Piazza della Scienza 3, 20126 Milano, Italy\label{aff175}
\and
School of Physical Sciences, The Open University, Milton Keynes, MK7 6AA, UK\label{aff176}
\and
Department of Astrophysics, University of Zurich, Winterthurerstrasse 190, 8057 Zurich, Switzerland\label{aff177}
\and
Theoretical astrophysics, Department of Physics and Astronomy, Uppsala University, Box 524, 751 20 Uppsala, Sweden\label{aff178}
\and
Mathematical Institute, University of Leiden, Einsteinweg 55, 2333 CA Leiden, The Netherlands\label{aff179}
\and
Univ. Lille, CNRS, Centrale Lille, UMR 9189 CRIStAL, 59000 Lille, France\label{aff180}
\and
Kobayashi-Maskawa Institute for the Origin of Particles and the Universe, Nagoya University, Chikusa-ku, Nagoya, 464-8602, Japan\label{aff181}
\and
Institute for Particle Physics and Astrophysics, Dept. of Physics, ETH Zurich, Wolfgang-Pauli-Strasse 27, 8093 Zurich, Switzerland\label{aff182}
\and
Department of Astrophysical Sciences, Peyton Hall, Princeton University, Princeton, NJ 08544, USA\label{aff183}
\and
Fakult\"at f\"ur Physik, Universit\"at Bielefeld, Postfach 100131, 33501 Bielefeld, Germany\label{aff184}
\and
Space physics and astronomy research unit, University of Oulu, Pentti Kaiteran katu 1, FI-90014 Oulu, Finland\label{aff185}
\and
American Museum of Natural History, Department of Astrophysics, New York, NY 10024, USA\label{aff186}
\and
SRON Netherlands Institute for Space Research, Landleven 12, 9747 AD, Groningen, The Netherlands\label{aff187}}    
   
%
%
%
%


%
%

\abstract{
Strong gravitational lens modelling requires the lens light, lensed source light, and contaminants to be separated in each image before mass modelling codes can fit the system. Within \Euclid this separation has so far been done by hand, which does not scale to the more than 15\,000 candidates in the Euclid Lens Identification Pipeline and Scientific Exploration (ECLIPSE) catalogue of Data Release 1 (DR1). We present \texttt{AstroVink-TREX}, a neural network that performs this separation automatically and predicts three flux maps, in a standard format directly usable by mass-modelling pipelines. The network uses a hybrid architecture, pairing a vision transformer encoder reused from the \texttt{AstroVink} lens detection network with a parallel convolutional branch. A custom loss function penalises both flux and structural error, while allowing independent weighting per component. From the network's predictions we also derive the modelling centre, fitting region, conjugate points, and binary maps, providing all the inputs that a modelling pipeline needs to run without manual setup. We evaluate the efficiency of the network on DR1 systems against existing \texttt{PyAutoLens} fits and through expert validation. Across approximately 3000 inspected systems, taking 75\% agreement between inspectors, just 39 lens models fail because of a segmentation error. Following this validation, the network provides the component separation and modelling setup for every system modelled in DR1. With automated segmentation, lens modelling is no longer limited by manual preprocessing and can now operate at the same scale as lens discovery.}

%

%

\titlerunning{Automated strong lens segmentation in DR1}
\authorrunning{Euclid Collaboration: Vincken et al.}
%
%

   \maketitle
%
%
%
%
\section{\label{sc:Intro}Introduction}
The \Euclid Space Telescope \citep{EuclidSkyOverview} is transforming the landscape of strong gravitational lens science through its combination of wide-area coverage, high spatial resolution, and depth. The Quick Data Release 1 (Q1) Strong Lensing Discovery Engine (SLDE) catalogue contains over 500 high-confidence strong lens candidates \citep{Q1-SP048, Q1-Ecker}, and for DR1 \citep{DR1cite} the ECLIPSE catalogue contains over 15\,000 candidates \citep{DR1-SL01}.

Modelling these systems at scale requires automated mass modelling pipelines such as \texttt{PyAutoLens} \citep{Nightingale2018, Nightingale2021, DR1-SL03}, \texttt{Lenstronomy} \citep{Birrer_2018, Birrer_2021}, \texttt{GLEE} \citep{Suyu2010, Suyu2012}, \texttt{Herculens} \citep{Galan2022}, \texttt{GIGA-Lens} \citep{Gu_2022}, and \texttt{pronto} \citep{vegetti_2009, rybak_2015, rizzo_2018, powell_2021, ndiritu_2024}. These models provide constraints on the mass distribution of the lens \citep[e.g.,][]{Gavazzi_2007, Nightingale_2019, Shajib_2024}, allow observations of magnified distant sources \citep[e.g.,][]{Welch2022}, and measure cosmological parameters including the Hubble constant \citep[e.g.,][]{Wong2019, TDCOSMO2025} and properties of dark matter \citep[e.g.,][]{vegetti2023stronggravitationallensingprobe, li2024}.

Before any of these codes can model a system, the lens and source light must be identified in the image, and contaminants such as  line-of-sight galaxies and imaging artefacts must be masked out. Automated solutions for this exist outside of \Euclid, such as \texttt{DOLPHIN} \citep{shajib_2025}, which uses a neural network (NN) for component recognition as part of its forward modelling pipeline with \texttt{Lenstronomy} applied to \textit{Hubble} Space Telescope (HST) data. Object detection tools such as \texttt{SExtractor} \citep{Bertin1996} and \texttt{SourceXtractor++} \citep{SourceXtractorA, SourceXtractorB} detect and catalogue objects in survey images but are not built for decomposing a lens system into its components. Within \Euclid, identifying and masking the different components of strong lens systems has so far been done entirely by hand: an expert maps contaminants and outlines the lens and source regions for each system, one at a time. This is slow, difficult to reproduce, and dependent on individual judgement. With the volume of lens candidates in our catalogue, this manual approach is no longer an option.

In this work we present \texttt{AstroVink-TREX} (Transformer for Ring Extraction), a NN that automates strong lens component separation. The outputs of the network are used for the full ECLIPSE catalogue, providing automated setup for all systems modelled in DR1 \citep{DR1-SL03}.

The network builds on \texttt{AstroVink}, a vision transformer (ViT) based strong lens detection network applied to \Euclid Q1 and DR1 data \citep{Q1-Vincken}. \texttt{AstroVink-TREX} reuses this pre-trained network and adds components that predict three continuous flux maps: one each for the lens, one for the lensed source, and a third one for nearby contaminants that should be masked before modelling. Additionally, from these flux maps, we automatically determine the relevant modelling region and centre, identify conjugate points, and create binary maps for each component. All segmentation outputs of the network are provided in standard Flexible Image Transport System (FITS) format that many mass modelling codes can use directly.

Segmentation requires training data in which each of the components is labelled separately, and no such data exists within \Euclid. We construct a custom training set by deriving targets from systems that have already been successfully modelled with \texttt{PyAutoLens}, which limits us to just a few hundred systems. A set this small would normally be far too limited to train a deep NN from scratch. However, by using \texttt{AstroVink}, we already have a network that recognises lens and source structure from its training for detection. The labelled set only has to teach it to separate that structure into distinct maps, rather than identify lensing features from scratch.

We evaluate the network on new strong lens candidates from DR1. Standard segmentation metrics measure how closely the predicted maps match the lens and source light from existing \texttt{PyAutoLens} models, which partly reflects their suitability for modelling but does not fully capture it. We therefore also feed the automated maps into the \texttt{PyAutoLens} pipeline and assess the resulting model fits, supported by visual inspection from strong lens modelling experts.

This paper is structured as follows. In Sect.~\ref{sc:Data} we describe the training data construction and the DR1 test set. Section~\ref{sc:Method} presents the network architecture, loss function, and training procedure. Results on performance validation are given in Sect.~\ref{sc:Results}, including the expert inspection campaign. Finally, Sect.~\ref{sc:Conclusion} concludes and discusses the application.

\section{\label{sc:Data}Data}
Building a segmentation network for strong lenses requires training data that includes target flux maps per component at pixel-level accuracy. For \Euclid strong lenses, no such data set exists. In this work we therefore created a custom training data set. The base for this data set comes from successfully modeled strong lens systems inside Q1. Specifically, we used 336 models from the main SLDE catalogue \citep{Q1-SP048} and an additional 41 from SLDE-F \citep{Q1-Ecker}, totalling 377 systems. These models were produced using the \texttt{PyAutoLens} \citep{Nightingale2018, Nightingale2021} modelling code.

For each object, we used two sources of data. From the SLDE catalogue we took cutouts of the \IE flux captured by the optical VIS instrument \citep{EuclidSkyVIS}, along with the root mean square (RMS) noise map and point-spread function (PSF). These originate from the VIS processing pipeline \citep{Q1-TP002, DR1-TP002} and are delivered through the \Euclid MERge Processing Function pipeline \citep{Q1-TP004, DR1-TP009}. From the \texttt{PyAutoLens} modelling outputs, we used the predicted lens light and the predicted source light as it appears in the image plane. Colour composites shown in this work additionally use \JE-band imaging from the Near Infrared Spectrometer and Photometer (NISP; \citealp{EuclidSkyNISP}) instrument.

\subsection{\label{sc:Signal}Signal extraction}
The \IE flux contains both signal and noise. To separate these, a per-pixel signal-to-noise ratio (S/N) is calculated using

\begin{equation}
\text{S/N} = \frac{\text{\IE}}{\text{RMS}} ,
\end{equation}
where $\text{\IE}$ is the observed flux per pixel and $\text{RMS}$ is the noise level per pixel. 

Noise produces both positive and negative S/N values, while real signal contributes positive values. We estimate the noise distribution by mirroring the negative S/N values around zero, giving a symmetric distribution with mean $\mu = 0$ and standard deviation $\sigma$. Here, $\mu$ represents the centre of the noise distribution, not the background surface brightness. We classify any pixel above a threshold $t$ as real signal. This threshold is defined using

\begin{equation}
t = \mu + n_t\,\sigma ,
\end{equation}
where $n_t$ is the number of standard deviations above the mean. The choice of $n_t$ impacts the balance between how much noise is kept and how many faint signal pixels are missed.

Instead of defining a constant number of standard deviations as is done in other work \citep[e.g.,][]{Bertin1996}, we determine a dynamic per-image $n_t$ using the Benjamini--Hochberg false discovery rate (FDR) method \citep{BenjaminiHochberg1995}. This method was applied to radio telescope imaging by \cite{Hopkins_2002} to address limitations of using a fixed $n\sigma$ threshold. One of these limitations relevant to this work is that a fixed threshold cannot always account for varying signal distributions across images, for example mixing up faint source images with noise pixels.

FDR determines $n_t$ separately for each image, rather than using a fixed number of standard deviations. The parameter $\alpha$ controls the fraction of noise pixels allowed among those classified as signal. We set $\alpha = 0.1$ (allowing 10\% noise contamination), to ensure preservation of very faint sources. Appendix~\ref{fig:fdr-threshold-mosaic} visualises the signal extraction approach for three representative lens systems in our data set.

\subsection{\label{sc:Components}Component separation}
The lens light and source light outputs from \texttt{PyAutoLens} are the best-fit model reconstructions of the lens galaxy and the lensed background source in the image plane. However, these model reconstructions represent the predicted flux from fitted light profiles, not the observed pixel values, so they do not perfectly match the flux in the original image. Training a network to predict the model outputs directly would therefore force it to reproduce pixel values that are not present in its input. To address this, we do not use the model outputs directly for training. Instead, they are used only to define spatial maps that indicate the expected location of each component. These maps are then applied to the thresholded signal image constructed in the previous step.

The lens component, $T_{\text{lens}}$, is the intersection between the signal image and the lens mask. The source component, $T_{\text{source}}$, is the intersection between the signal image and the source mask. The remaining pixels are assigned to the contaminant component, $T_{\text{cont}}$, which includes unrelated line-of-sight galaxies and any residual flux that should be masked when creating a lens model. The result is that for each system, we have the original input image paired with three target maps representing the lens, source, and contaminant flux.

\subsection{Training set and preprocessing}
\begin{figure}
    \centering
    \includegraphics[width=\linewidth]{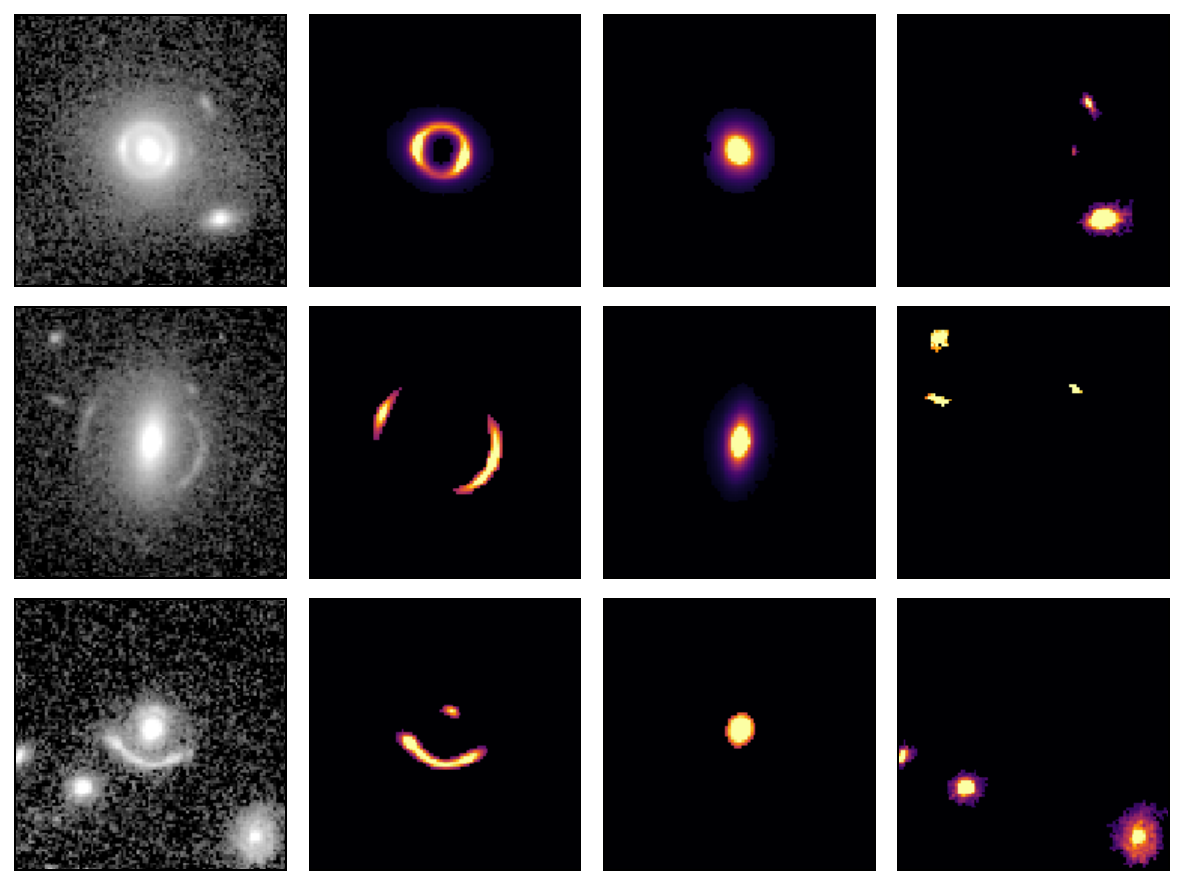}
    \caption{Example objects from the training data set. With the input image (column one), the target source map (column two), the target lens map (column three) and the target contaminants map (column four). Target maps are displayed using a colour gradient for visualisation purposes; during training the targets are used in grey-scale.}
    \label{fig:training-examples}
\end{figure}

After producing the target maps, the 377 training samples were randomly split into 302 (80\%) training and 75 (20\%) validation samples. For each system, the input is the \IE image with an arcsinh stretch applied to compress the dynamic range, replicated across three channels to form an RGB image. The image was then resized to $256 \times 256$ pixels and normalised using the standard \texttt{ImageNet} channel means \((0.485, 0.456, 0.406)\) and standard deviations \((0.229, 0.224, 0.225)\). For each channel, the mean was subtracted and the result divided by the corresponding standard deviation.
We acknowledge that the processing may introduce a small shift in pixel brightness, though it preserves the spatial structure. We accept this because the frozen \texttt{AstroVink} encoder was pre-trained on inputs with this same distribution, so applying the identical transformations here keeps its learned feature representations valid. Examples of training samples are shown in Fig~\ref{fig:training-examples}.

\subsection{\label{sc:test-set}DR1 data and test set}
While the training sets were constructed from existing Q1 data, the evaluation of the network was performed on new and unseen data. For this we used strong lens candidates from the ECLIPSE catalogue \citep{DR1-SL01} of DR1. This catalogue was assembled using machine-learning models \citep{DR1-Vincken-a}, followed by a visual inspection pipeline \citep{DR1-SL02}, and lens models created using \texttt{PyAutoLens} \citep{DR1-SL03}.

We used a preliminary subset of 750 modelled DR1 lens candidates to test general performance and the impact of training choices such as loss function. In contrast to the training set, where the \texttt{PyAutoLens} outputs are used only to define spatial maps, the test set uses the reconstructed lens and source light from the lens models to compare network predictions.

The \texttt{PyAutoLens} models reconstruct only the lens and source light, so no reference exists for the contaminant channel. Test-set metrics are therefore restricted to the lens and source channels, and the contaminant channel is assessed on the validation set (Sect.~\ref{sc:BestModel}) and through the expert campaign (Sect.~\ref{sc:Experts}), where a mismasked contaminant is recorded directly as a segmentation failure.

\section{\label{sc:Method}Method}
Our segmentation method is built on a NN architecture following an encoder and decoder structure. The encoder is responsible for making a summary of features it sees in the image, while the decoder interprets that summary to predict the output maps. In our case, the output is three separate flux maps for the lens, source, and contaminant components. To create a successful network for strong lens component separation, our architecture has to satisfy the following criteria.

\begin{enumerate}[label=(\roman*)]
    \item Predict each component's flux accurately, since the lens and source flux may overlap and a pixel can hold both.
    \item Resolve the boundaries between components, since a missed source image or a contaminant counted as source flux can bias the mass model.
    \item Connect parts of the image that are far apart, since the arcs of one source are often disconnected on the sky, yet belong to the same component.
    \item Preserve fine spatial detail through the network, since small-scale structure such as thin or faint arcs can be lost when the image is compressed during encoding.
    \item Optimise from a small data set, since only a few hundred labelled systems are available.
\end{enumerate}

No standard architecture immediately covers all of these criteria. Convolutional neural networks (CNNs) work by sliding small filters across the image, and checking at every position how well the pixels there match a pattern the network has learned. This allows learning fine detail, but makes them weaker at relating features that are far apart in an image. ViTs handle this through self-attention, where the image is split into a grid of small square pixel blocks called patches, with each patch being compared against every other to give a global view. However, this comes at the cost of fine detail, since the network sees each patch as a whole rather than the pixels inside it. 

We therefore use a hybrid network architecture, combining a ViT with CNN layers so that each makes up for the other's weakness. The transformer gives the global view that connects distant structure, and the convolutional layers keep the fine local detail.

This hybrid network outputs three continuous flux maps, one per component, satisfying criterion (i). This is done by using convolutional layers to detect local structure such as the boundaries between objects (ii), while the ViT encoder's self-attention connects distant parts of the image that belong to the same source (iii). The encoder's patch-level output is passed through a convolutional decoder, which is a set of convolutional layers that turns the output back into a full-size image. In parallel, a second convolutional branch reads the original image directly to preserve pixel-level detail at full resolution (iv). Since the \texttt{AstroVink} encoder \citep{Q1-Vincken} is already trained on a large set of \Euclid lenses, it brings an understanding of lens and source morphology that we could not learn from a few hundred systems alone (v). Finally, the information from both branches is brought together at each pixel and used to produce the three flux maps.

\subsection{\label{sc:Loss}Loss function}
Standard segmentation losses such as cross-entropy \citep{crossentropy1999} or Dice loss \citep{Milletari2016} treat segmentation as per-pixel classification, assigning each pixel to exactly one class. This approach fails for strong lens segmentation because lens and source light physically overlap in the image, requiring continuous flux prediction for each channel independently rather than classification.

A pure flux regression loss would treat all pixels equally, but not all errors are equally costly when outputs are used for lens modelling. A small missing image from the source can make the entire mass model fail, while tiny flux differences are more tolerable. To address this, we design a custom loss function that penalizes under-prediction on signal pixels more heavily than over-prediction. Additionally it provides weights for balancing structural versus brightness errors, and individual channel importance. The total loss is

\begin{equation}
L_{\text{total}} = \sum_{c} \left( w_c \, L_{\text{structure},c} + \lambda \, L_{\text{flux},c} \right) ,
\end{equation}
where we sum over channels $c$ (lens, source, contaminant), $w_c$ is the per-channel weight, and $\lambda$ scales the relative importance between the structure ($L_{\text{structure}}$) and flux ($L_{\text{flux}}$) terms. $L_{\text{structure}}$ for channel $c$ penalizes missing and invented signal using

\begin{align}
L_{\text{structure},c} ={}&
\underbrace{
\frac{1}{|S_c|}
\sum_{x \in S_c}
\max\left[0, T_c(x) - P_c(x)\right]
}_{\text{missing signal}}
\nonumber\\
&+
\underbrace{
\frac{1}{|\bar{S}_c|}
\sum_{x \in \bar{S}_c}
\max\left[0, P_c(x)\right]
}_{\text{invented signal}} \;,
\end{align}
where $S_c$ is the set of pixels where channel $c$ has signal, $\bar{S}_c$ is the set of background pixels, $T_c(x)$ is the target flux at pixel $x$, and $P_c(x)$ is the predicted flux at pixel $x$.

For $L_{\text{flux}}$ per channel $c$, we obtain the mean absolute error between predicted and target flux using

\begin{equation}
\label{eq:flux_loss}
L_{\text{flux},c} = \frac{1}{|S_c|} \sum_{x \in S_c} \left| P_c(x) - T_c(x) \right|,
\end{equation}
where the absolute difference between predicted and target flux is averaged over all signal pixels in $S_c$. We apply this only to signal pixels so that the large number of background pixels does not dominate the loss.

While building the loss function, we verified the contribution of each term by training and testing three models. One using only $L_{\text{structure}}$, another using just $L_{\text{flux}}$, and a final one using $L_{\text{total}}$. For $L_{\text{total}}$, we set $\lambda = 1.0$ as a starting point. We later tested more values for $\lambda$ to find the best balance between terms. Results on these experiments are described in Sect.~\ref{sc:LossExperiment}.

For the per-channel weights, we measured signal fraction (pixels above background) and mean flux over those pixels for that component across the training set. All fluxes are given in ADU\,s$^{-1}$ per pixel. The lens has 1.2\% coverage with mean flux 0.25, source has 2.3\% with mean flux 0.03, and contaminants have 4.1\% with mean flux 0.03. To find what dominates the loss, we took the product of these values, giving $3.2 \times 10^{-3}$ for lens, $7.2 \times 10^{-4}$ for source, and $1.1 \times 10^{-3}$ for contaminants. We weighted inversely proportional to these values to equalize each channel's contribution to the loss. This gives $w_{\text{lens}} = 1.0$, $w_{\text{source}} = 4.4$, and $w_{\text{cont}} = 3.0$ as the channel weights we adopt for our loss function.

\subsection{\label{sc:Metrics}Performance metrics}
The loss function optimises both structural and flux accuracy, so we evaluated the network against both. Besides total and per-channel loss, we tracked two performance metrics during training, validation, and testing. We did not use a binary-overlap metric such as Intersection over Union, since our channels may overlap and have continuous flux rather than exclusive class labels.

The first metric is the mean absolute error (MAE) on support, which we denote using $E_{\mathrm{supp}}$. This is the standard MAE with a custom alteration that restricts to signal pixels only. Restricting it this way avoids a bias, since the large number of correctly predicted background pixels would otherwise dominate the metric. It is the same expression as the flux loss term $L_{\text{flux},c}$ (Eq.~\ref{eq:flux_loss}), so tracking it shows directly how well the network learns flux per channel.

The second metric is the Pearson correlation coefficient $r$. It measures whether the predicted and target flux follow the same spatial brightness pattern. A value of $r = 1$ means the prediction is bright wherever the target is bright and faint wherever the target is faint. A prediction that misses a source image or invents flux therefore scores poorly, so $r$ captures both structural and flux agreement in a single value. It is also insensitive to the absolute flux scale. This matters for the test set, where our predicted flux comes from the observed image while the target comes from the \texttt{PyAutoLens} model fit, so the two are not on the same absolute scale.

\begin{figure*}
    \centering
    \includegraphics[width=1\linewidth]{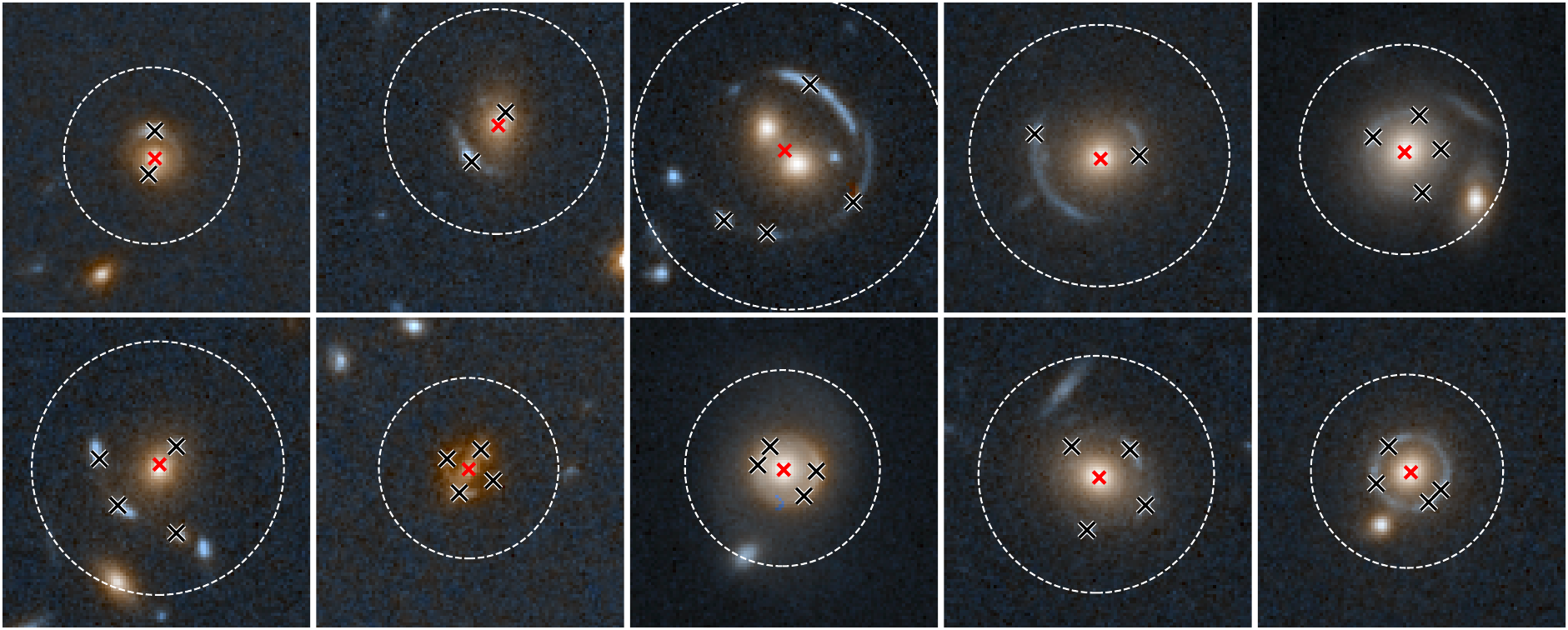}
    \caption{Derived segmentation outputs overlayed on ten DR1 lens systems. The red cross marks the centre of the modelling region, the dashed circle marks the modelling region, and the blue crosses mark the conjugate points. These examples illustrate the performance of the automated procedure across a random subset of lens configurations, which may include more complex morphologies such as group scale systems with multiple sources or double-source-plane lenses.}
    \label{fig:derived-inputs}
\end{figure*}

\subsection{\label{sc:Training}Training approach}
Training a segmentation network on only a few hundred samples poses a significant overfitting risk, meaning that the network memorises the training examples instead of learning patterns that generalise to new systems. This risk is particularly high with our encoder that contains millions of parameters pre-trained on a much larger data set. To mitigate this, we froze the encoder weights entirely and only optimised the decoder. This preserves the feature representations the \texttt{AstroVink} encoder learned during pre-training on over a hundred thousand \Euclid strong lens images, while the decoder learns to map these features to pixel-level flux predictions.

To further reduce overfitting, we applied four $90^{\circ}$ rotations and horizontal flips to each sample during training. These transformations only reorder pixels and leave their values unchanged. We applied the same transform to the input image and all three target maps.

We optimised the network using AdamW \citep{kingma2017adammethodstochasticoptimization, loshchilov2019decoupledweightdecayregularization}, an algorithm that repeatedly adjusts the network weights to reduce the loss. The size of these adjustments is set by the learning rate, which started at $10^{-4}$ and followed a cosine annealing schedule that smoothly decreased it over training. In this way the network makes large adjustments early on and smaller, more precise updates as it converges. A weight decay of $10^{-3}$ penalises large weights, discouraging the network from relying on any single feature. 

We set the batch size to $16$, meaning the loss is averaged over $16$ systems before each update. This follows \citet{Masters2018RevisitingSB}, who show that small batch sizes between $2$ and $32$ consistently outperform larger batches by providing more frequent gradient updates with higher variance.

Finally, training was performed on a single NVIDIA RTX A4500 GPU using mixed precision \citep{micikevicius2018mixed} to reduce memory usage. We then applied early stopping with a patience of 50 epochs on the total validation loss, meaning training automatically halts if the loss does not improve for that number of consecutive epochs. 

\subsection{\label{sc:ModelInputs}Derived modelling inputs}
The main output of \texttt{AstroVink-TREX} is the set of predicted flux maps for each component. However, they also provide several secondary applications which are useful to initialise automated lens modelling.

One of these applications is that we use the predicted flux of the lens component to define the centre of the system. We take all pixels with predicted lens flux above zero and use the mean of their pixel coordinates as the system centre. We use this rather than the arbitrary cutout centre because the system may be offset from the centre of the cutout. The radius is then set by measuring the Euclidean distance from this centre to every pixel with predicted source flux above zero, and taking the largest of these distances. We then add a fixed \ang{;;2.0} margin. This margin simply provides a small buffer beyond the predicted source extent, so the modelling region is less sensitive to small inaccuracies in the predicted source boundary or system centre.

The same source flux can also be used to estimate conjugate points for modelling codes that use multiple-image positions as constraints. These are obtained by identifying bright, spatially distinct local maxima in the predicted source flux map and converting their pixel positions to angular coordinates. Appendix~\ref{fig:model-input-workflow} shows a worked example of the full process, from the \texttt{AstroVink-TREX} input image and predicted flux maps to the derived quantities used by the modelling pipeline.

Figure~\ref{fig:derived-inputs} shows the derived quantities for a set of DR1 systems. All three were used to initialise the modelling for the full catalogue, replacing the default definitions of the modelling centre, region, and multiple-image positions.

\section{\label{sc:Results}Results}
We evaluate \texttt{AstroVink-TREX} in two stages. First, we used the DR1 test set to tune the loss function and assess flux correlation between model fits. Second, we assessed the network's practical impact through the expert validation campaign, in which lens modellers inspected another set of models built from the network's outputs.

\subsection{\label{sc:LossExperiment}Loss-function experiment}
\begin{figure*}
    \centering
    \includegraphics[width=\linewidth]{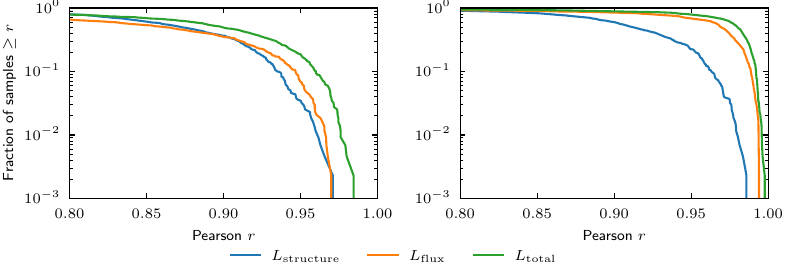}
    \caption{Distribution of Pearson $r$, on the test set for three loss configurations on the source (left) and lens (right) channels. The $x$-axis shows Pearson $r$ and the $y$-axis shows the fraction of samples achieving at least that correlation. With $L_{\text{structure}}$ (blue) using the structure-only term, $L_{\text{flux}}$ (orange) using the flux-only term, and $L_{\text{total}}$ (green) using the combination of both.}
    \label{fig:loss_ablation}
\end{figure*}

In the custom loss function, before tuning the relative weight between structure and flux, we first tested whether each term is needed. Figure~\ref{fig:loss_ablation} shows the distribution of Pearson $r$ between predicted and target flux across the test set for three models trained with different loss terms. For both channels the $L_{\text{flux}}$ term outperforms $L_{\text{structure}}$. However, $L_{\text{total}}$ outperforms individual terms in both channels, confirming that the two terms address complementary failure modes that neither can handle alone.

\begin{figure*}
    \centering
    \includegraphics[width=\linewidth]{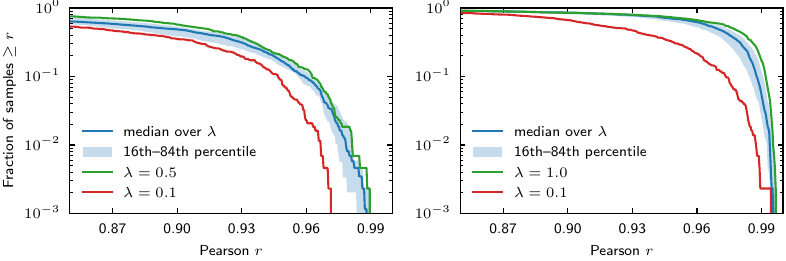}
    \caption{Pearson correlation $r$ between the predicted lens and source fluxes of \texttt{AstroVink-TREX}, and the lens and source light from the \texttt{PyAutoLens} models, evaluated on the DR1 test set for the source (left) and lens (right) channels. The $x$-axis shows $r$ and the $y$-axis the fraction of test samples reaching at least that value. The blue line is the median over the ten values of $\lambda$ and the shaded band the 16th--84th percentile across them. The green and red lines show the best and worst individual $\lambda$ runs.}
    \label{fig:lam_pearson}
\end{figure*}

Besides measuring the performance on the individual loss terms, we trained ten versions of the segmentation network to test which $\lambda$ gives the best balance between structural and flux accuracy. Since $\lambda$ controls how much weight the flux term receives relative to the structure term in the loss function, low values prioritize getting the right pixels while high values prioritize getting the right brightness. We tested values from $\lambda = 0.1$ to $\lambda = 5.0$ and evaluated each on our test set by measuring the Pearson correlation between the predicted flux and the \texttt{PyAutoLens} model fits. The results of this test are presented in Fig.~\ref{fig:lam_pearson}, where we show the best and worst performing $\lambda$ alongside the median and spread across all ten. The median $r$ for every individual $\lambda$ is listed in Appendix~\ref{table:lam_results}.

\begin{figure}
    \centering
    \includegraphics[width=\linewidth]{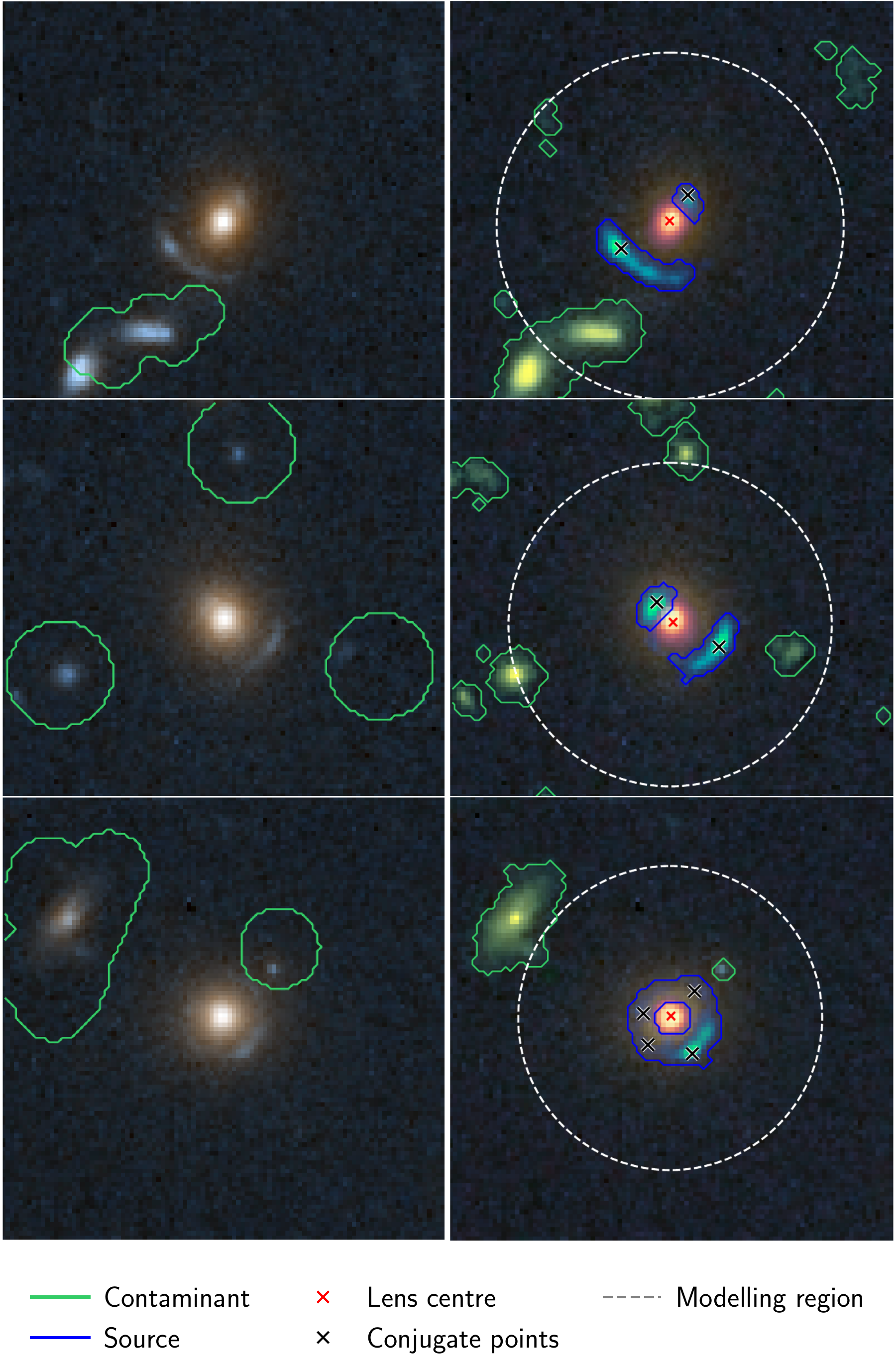}
    \caption{Comparison of manually drawn contaminant masks and the automated segmentation from \texttt{AstroVink-TREX} for three strong lens systems. Each row shows one system, with the manually drawn mask on the left and the automated prediction on the right. The green outlines indicate the contaminant masks, while the right panels additionally show the predicted lens and source light, their binary outline of the predicted source, the estimated lens centre, conjugate points, and relevant modelling region. All panels use colour cutouts constructed from \IE, \YE, and \JE with MTF scaling for visualisation.}
    \label{fig:manual-auto-segmentation}
\end{figure}

Lens flux correlation improves with $\lambda$ and plateaus at $\lambda \geq 1.0$, while source correlation peaks at $\lambda = 0.5$ and drops at higher values as the network begins to miss faint arcs without enough structural guidance. The source flux correlation also shows increasing variance across samples at higher $\lambda$, meaning predictions become less reliable when flux loss dominates. We adopt $L_{\text{total}}$ with $\lambda = 0.5$, since it provides the most stable performance across both measurable channels.

\subsection{\label{sc:BestModel}Best model performance}
The final \texttt{AstroVink-TREX} network was saved using early stopping by the lowest validation loss, while training itself was allowed to continue until no further improvement was found. The final best model was saved at epoch 175, and training stopped at epoch 225. With an epoch time of approximately 7 seconds on the small data set, the full training run took about 30 minutes.

The best model reached a validation loss of approximately 0.1 on $L_{\text{total}}$. The lens, source, and contaminant channels contributed 0.04, 0.04, and 0.02 to the total loss, respectively. These are the weighted contributions defined in Sect.~\ref{sc:Loss}. The lens and source contribute equally despite the lens being the brighter component, showing that the per-channel weighting achieves the intended balance. The contaminant channel contributes least, indicating that this channel could potentially benefit from more tuning in later stages. 

As an additional check, we tracked the $E_{\mathrm{supp}}$ as defined in Sect.~\ref{sc:Metrics}. This gives errors of 0.02 for the lens, 0.01 for the source, and 0.01 for the contaminant channel, showing that the predicted flux is close to the target flux in the regions where each component is present.

\subsection{\label{sc:Experts}Expert validation campaign}
The results so far measure the segmentation maps against reference maps, not their effect when used for modelling. Whether a segmentation map produced by a NN is good enough to model a lens can only be judged from the resulting model, so a dedicated visual inspection campaign was carried out. 

In the inspection, a group of expert lens modellers reviewed \texttt{PyAutoLens} models created using the products of \texttt{AstroVink-TREX}. This inspection set consists of approximately 3000 systems and is part of the final catalogue. Each model was classified against a fixed set of criteria. Any model that failed was additionally assigned the pipeline stage responsible for the failure: the segmentation; the foreground lens-light subtraction; or the mass and source reconstruction. The full criteria, inspection details, and validation results on the modelling itself are described in \citet{DR1-SL03}. In this work we analyse the validation results specifically related to the impact of using automated segmentation.

Across the inspected sample, the segmentation causes a low number of failures. A classification is accepted only when a chosen fraction of inspectors agree on it. Following \citet{DR1-SL03}, we adopted a consensus threshold of 75\%. At this level, 210 systems (7.0\% of the sample) are marked as failures, and of these the segmentation is responsible for 18.6\%. In absolute terms these are 39 systems of all inspected models for which a segmentation error is the cause of a failed fit. This ordering is stable across thresholds. Segmentation accounts for 0.4\% of all inspected systems at 100\% consensus, 1.3\% at 75\%, and 3.4\% at 50\%. It is important to note that these percentages reflect segmentation errors that caused the lens model to fail, rather than all segmentation errors. Errors that did not significantly affect the resulting mass model may therefore not have been classified as failures.

\begin{figure*}
    \centering
    \includegraphics[width=0.72\linewidth]{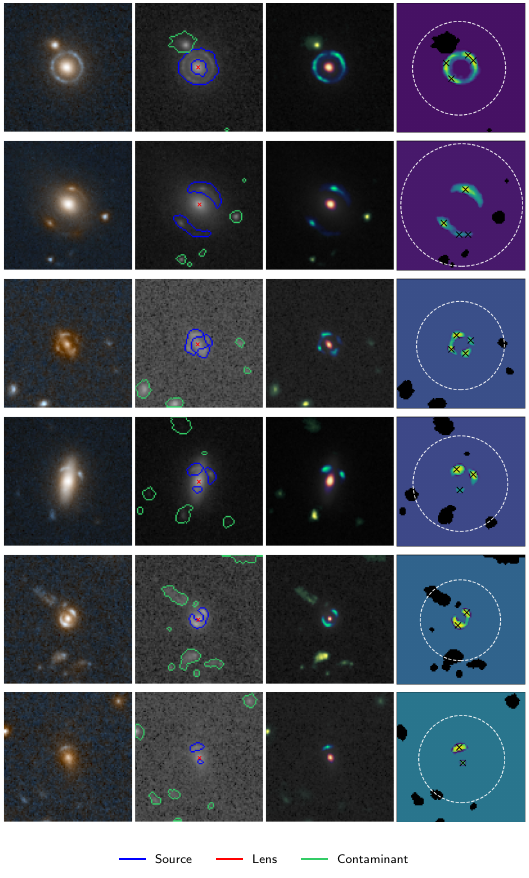}
    \caption{Final network predictions and their application to lens modelling, showing examples of successful segmentation on \Euclid DR1 systems with each row showing one system in four panels. This shows the observed system (panel one) in \IE+\JE in MTF scaling for visualisation purposes, the binary outline for each segmented map (panel two) based on predicted flux (panel three). Finally we show the way that each segmentation output is used as input for the modelling pipeline (panel four), with the segmentation maps used to detect conjugate points, define the radius for the relevant region, and mask to remove nearby contaminants.}
    \label{fig:segmentation-success}
\end{figure*}

\begin{figure*}
    \centering
    \includegraphics[width=0.72\linewidth]{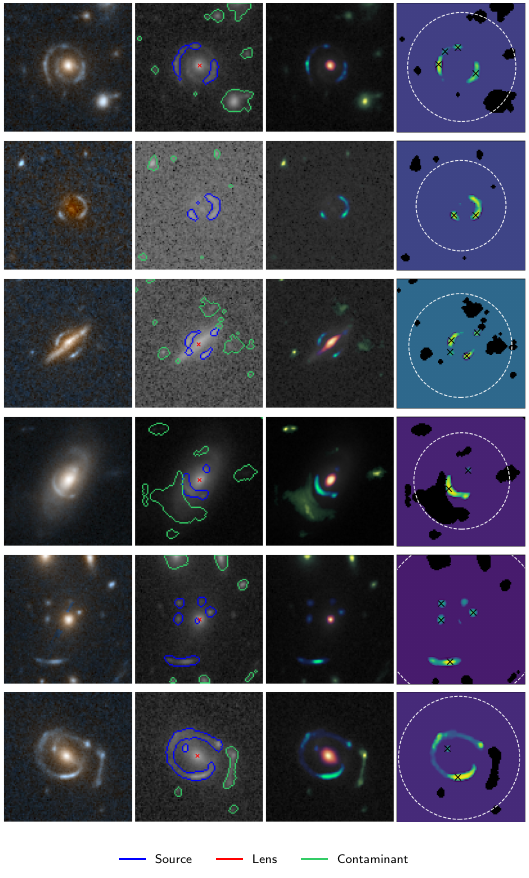}
    \caption{Final network predictions and their application to lens modelling, showing examples of common failure modes on segmented DR1 systems as defined by expert validation, with each row showing one system in four panels. Displaying the observed system (panel one) in \IE+\JE in MTF scaling for visualisation purposes, the binary outline for each segmented map (panel two) based on predicted flux (panel three). Finally we show the way each segmentation output is used as input for the modelling pipeline (panel four), with the segmentation maps used to detect conjugate points, define the radius for the relevant region, and mask to remove nearby contaminants. The first two rows display examples of a missing source or lens, the next two rows display contaminant confusion with lens light, and the final two rows show more complex morphologies that display the confusion with (potential) source images.}
    \label{fig:segmentation-failures}
\end{figure*}

Figure~\ref{fig:segmentation-success} shows examples of success cases. The predicted maps separate the lens, source, and contaminants. From these maps we derive the conjugate points, the fitting radius, and the contaminant mask that the modelling pipeline requires. Figure~\ref{fig:segmentation-failures} shows the failures that did occur. In each case the mask misplaced flux in a way that damaged the fit, such as a lensed image removed as though it were a line-of-sight galaxy, a contaminant left only partly masked, or nearby galaxies missed entirely.

\section{\label{sc:Conclusion}Conclusions}
In this work we have presented \texttt{AstroVink-TREX}, a machine-learning solution that automates the separation of strong lens systems into their lens, source, and contaminant components. The automated segmentation removes much of the manual preparation previously required for lens modelling. Figure~\ref{fig:manual-auto-segmentation} gives a final comparison between manually drawn contaminant masks and the predictions from \texttt{AstroVink-TREX} for three strong lens systems. The automated masks identify the main contaminant regions without requiring them to be drawn individually. The figure also shows how the predicted lens and source light are used to estimate the lens centre, conjugate points, and relevant modelling region. Unlike previous \Euclid modelling, which used a fixed-size region, the automated approach determines the modelling region separately for each system.

The reduction in manual work is particularly relevant for the more than 15\,000 candidates in the \Euclid DR1 ECLIPSE catalogue. An estimate of the time required for manual preprocessing comes from the COSMOS-Web Lens Survey (COWLS; \citealp{Nightingale_2025}). For this sample, 419 systems required manually drawn contaminant masks and selected conjugate points. The modeller estimates that these tasks require at least one hour per 50 systems. At this rate, preparing 15\,000 systems would require approximately 300 hours of manual work. In comparison, we benchmarked \texttt{AstroVink-TREX} on 50 systems using a standard laptop CPU without GPU acceleration or parallel processing across systems. The average network inference time was 0.34 seconds per system, increasing to 0.35 seconds when including image preprocessing and downstream processing such as conjugate point detection. When scaling this measurement to 15\,000 systems, this gives approximately 1.5 hours of automated processing. 

Expert validation of the resulting \texttt{PyAutoLens} models shows that the segmentation performance is reliable enough to model lenses across the full catalogue. The network handles the general population of lenses, and the systems it cannot segment correctly are a small and identifiable set rather than a limitation across the whole sample. 

The current limitations of the network mainly involve confusion between source structures and contaminants. This happens particularly when source images are very faint or the system has a complex morphology, making the different components difficult to distinguish. These failures can be identified automatically by comparing the lens and source light reconstructed by the mass modelling pipeline with the segmentation maps predicted by \texttt{AstroVink-TREX}, allowing systems with significant disagreement to be flagged for further inspection.

The successfully modelled systems provide a way to improve the network further. Each has a \texttt{PyAutoLens} model fit that can be used to derive new training targets, in the same way that the original training set was constructed. As new systems are modelled, the network can be retrained on an increasingly diverse sample and improve its generalisation to harder systems.

Besides the additional training samples, segmentation could be improved further by tuning the training procedure and loss function. In particular, the contaminant channel contributes less to the validation loss than the lens and source channels, indicating that this part of the network could benefit from additional optimisation. The current failure cases could also be used to identify the types of systems that are most difficult for the network and guide the selection of additional training examples.

One other limitation is that the network currently uses only one band, while colour could help separate the components. In this work we use only information from \IE because the frozen encoder was pre-trained on this single-band input, and adding the NIR bands would require adapting or retraining it with a larger data set. Besides refining the current \texttt{AstroVink-TREX} model, a next step would be to rebuild the full \texttt{AstroVink} architecture to allow multi-band inputs.

Finally, we emphasise that the network is not tied to \texttt{PyAutoLens}. Its outputs are predicted maps provided in the FITS format that mass modelling codes often read directly. The segmentation therefore does not need to be modified or reformatted for a different code. It is applicable to any mass-modelling pipeline that requires the lens, source, and contaminants to be separated before modelling. With this automated separation in place, strong lens modelling can be extended to the full samples expected from future \Euclid data releases.

%
%

\begin{acknowledgements}
\AckDRone
\AckEC  
We gratefully acknowledge the Zooniverse team for their invaluable work on the host platform for Space Warps. 
This project would not be possible without the immense contributions of citizen-scientists. 
Those contributing classifications for this project are listed on \href{https://www.zooniverse.org/projects/aprajita/space-warps-esa-euclid/about/team}{`The Team'} webpage.
We greatly appreciate their commitment, enthusiasm, and time.
This project has received funding from the European Research Council (ERC) under the European Union’s Horizon 2020 research and innovation programme (LensEra: grant agreement No 945536). TEC is funded by the Royal Society through a University Research Fellowship. 
"ELSA: Euclid Legacy Science Advanced analysis tools" (Grant Agreement no. 101135203) is funded by the European Union. Views and opinions expressed are however those of the author(s) only and do not necessarily reflect those of the European Union or Innovate UK. Neither the European Union nor the granting authority can be held responsible for them. UK participation is funded through the UK Horizon guarantee scheme under Innovate UK grant 10093177.
Part of this research was carried out at the Jet Propulsion Laboratory, California Institute of Technology, under a contract with the National Aeronautics and Space Administration (80NM0018D0004).
FC is supported by the Agencia Estatal de Investigaci\'on (AEI), Ministerio de Ciencia, Innovaci\'on y Universidades, Spain, under the project PID2024-155455NB-I00, within the 2024 Call for Knowledge Generation Project and by project CNS2023-144016, and co-funded by the European Union -- NextGenerationEU.
AS acknowledges support from the National Key R\&D Program of China (No. 2023YFA1607800, 2023YFA1607802), and from the Yangyang Development Fund. 
AN gratefully acknowledges the support of the Belgian Federal Science Policy Office (BELSPO) for the provision of financial support in the framework of the PRODEX Programme of the European Space Agency (ESA) under contract number 4000143347.
AG acknowledges support by the SNSF through PostDoc Mobility Return grant P5R5-2\_235360.
NL is supported through a graduate studentship from the UKRI STFC and the University of Portsmouth, and acknowledges travel support provided by Euclid:UK through STFC grant APP91126 and by STFC for UK participation in the Rubin LSST through grant ST/X001334/1.
PH is funded by the Royal Society (RF\textbackslash{}ERE\textbackslash{}210033) and acknowledges travel support provided by Euclid:UK through STFC grant APP91126.
WJRE has received funding from the European Research Council (ERC) under the European Union’s Horizon 2020 research and innovation programme
(LensEra: grant agreement No 945536). WJRE acknowledges the support by the Italian Ministry of University and Research (grant FIS2023-01611, CUP C53C25000300001).
Travel supported by Euclid:UK through STFC grant APP91126
S.K.L acknowledges the generous support from the Croucher Foundation through the Croucher Postdoctoral Fellowship.
QH was supported by the Dutch Research Council (NWO) through the grant Open Competitie ENW M21-1.
JWN is supported by an STFC/UKRI Ernest Rutherford Fellowship, Project Reference: ST/X003086/1.  
Numerical computations were done on the Sciama High Performance Compute
(HPC) cluster which is supported by the ICG, SEPNet and the University of Portsmouth. 
DJ acknowledges support by the First Rand Foundation, South Africa, and the Centre National de la Recherche Scientifique of France.
CJS acknowledges funding from the NASA Euclid General Investigator Program Grant No. 80NSSC25K7735
\end{acknowledgements}
%

\bibliography{Euclid, Q2, my} 
%
%

%

\begin{appendix}

\section{\label{sc:AppSignal}Signal extraction}
\begin{figure}[!t]
    \centering
    \includegraphics[width=\linewidth]{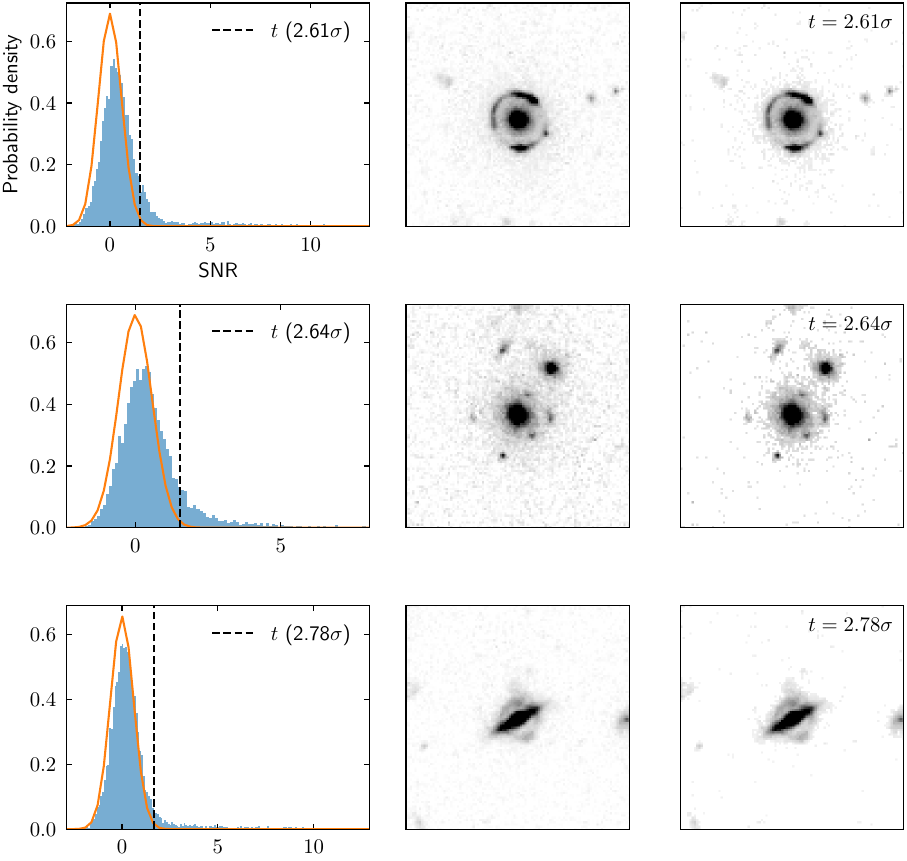}
    \caption{Signal extraction for three lens systems. The left panels show the S/N histograms with fitted Gaussian noise models (orange curves) and adaptive FDR thresholds (dashed lines). The central panels show the observed \IE flux. The right panels show the \IE flux after applying the threshold, with the corresponding threshold value $t$ indicated in each panel. All images are displayed in gray-scale with an intensity scaling defined by $v_{\max}$ (99th percentile of the raw flux) and $v_{\min}=0$.}
    \label{fig:fdr-threshold-mosaic}
\end{figure}

This appendix illustrates the FDR-based signal extraction described in Sect.~\ref{sc:Signal}. Figure~\ref{fig:fdr-threshold-mosaic} shows three representative lens systems, with the fitted noise model, the resulting per-image threshold, and its effect on the \IE flux.

\section{\label{sc:AppLam}Loss weight values}
This appendix reports the individual results of the $\lambda$ experiment described in Sect.~\ref{sc:LossExperiment}. Table~\ref{table:lam_results} lists the median Pearson $r$ for each tested value.

\begin{table}[!t]
    \centering
    \caption{
    Segmentation performance of the \texttt{AstroVink-TREX} network as a function of the loss weight $\lambda$, which sets the relative contribution of the structure and flux terms in the combined loss. The first column lists the value of $\lambda$. The second and third columns report the median Pearson correlation coefficient $r$ between the predicted flux and the \texttt{PyAutoLens} model fits across the DR1 test set, for the source channel and the lens channel, respectively. Higher values indicate closer agreement.}
    \smallskip
    \label{table:lam_results}
    \smallskip
    \renewcommand{\arraystretch}{1.15}
    \begin{tabular}{l@{\hskip 3em}c@{\hskip 3em}c}
        \hline\hline
        $\lambda$ & Source $r$ & Lens $r$ \\
        \hline
        0.1 & 0.863 & 0.927 \\
        0.2 & 0.902 & 0.964 \\
        0.3 & 0.906 & 0.961 \\
        0.5 & 0.913 & 0.970 \\
        0.7 & 0.906 & 0.969 \\
        1.0 & 0.900 & 0.977 \\
        1.5 & 0.871 & 0.977 \\
        2.0 & 0.871 & 0.970 \\
        3.0 & 0.891 & 0.970 \\
        5.0 & 0.891 & 0.967 \\
        \hline
    \end{tabular}
\end{table}

\begin{figure*}
    \centering
    \includegraphics[width=\linewidth]{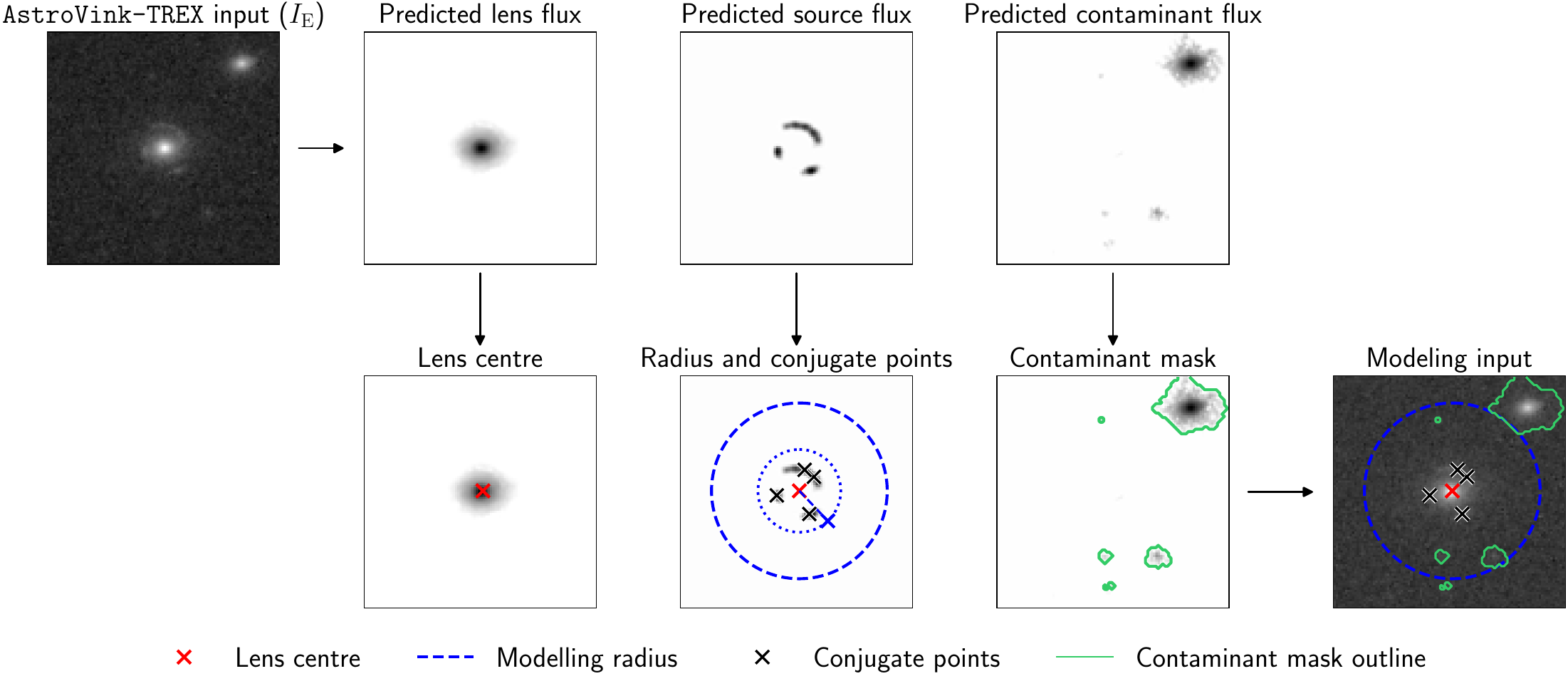}
    \caption{Worked example showing how the outputs of \texttt{AstroVink-TREX} are converted into inputs for lens modelling. From the \IE input image, the network predicts separate lens, source, and contaminant flux maps. The lens flux is used to determine the system centre, the source flux to determine the modelling radius and conjugate points, and the contaminant flux to construct the contaminant mask. These quantities are then combined into the inputs used to initialise the lens model.}
    \label{fig:model-input-workflow}
\end{figure*}

\end{appendix}

\label{LastPage}
\end{document}